\documentstyle[12pt]{article}

\newcommand{\be}{\begin{equation}}
\newcommand{\ee}{\end{equation}}
\newcommand{\bref}[1]{(\ref{#1})}
\newcommand{\rf}[1]{\,\ref{#1}}
\newcommand{\ct}[1]{\,\cite{#1}}

\def\NN{N\kern-.25em N}
\def\lcm{\mathop{\rm lcm}}
\def\ex#1{e^{#1\pi i / N}}
\def\Tr{{\rm{Tr}}}

\makeatletter

\def\secteqno{\@addtoreset{equation}{section}%
\def\theequation{\thesection.\arabic{equation}}}
\def\secteqnoA{\@addtoreset{equation}{section}%
\def\theequation{{\rm A}.\arabic{equation}}}
\def\endsecteqno{\def\theequation{\@ifundefined{chapter}%
{\arabic{equation}}{\thechapter.\arabic{equation}}}}
\newcounter{subequation}
\def\thesubequation{\alph{subequation}}
\def\sneqnarray{\stepcounter{equation}\let\@currentlabel=\theequation
\setcounter{subequation}{1}
\def\@eqnnum{{\rm (\theequation.\thesubequation)}}
\global\@eqcnt\z@\tabskip\@centering
\let\\=\@eqncr\let\@@eqncr=\@@sneqncr
$$\halign to \displaywidth\bgroup\@eqnsel\hskip\@centering
 $\displaystyle\tabskip\z@{##}$&\global\@eqcnt\@ne
 \hskip 2\arraycolsep \hfil${##}$\hfil
 &\global\@eqcnt\tw@ \hskip 2\arraycolsep
$\displaystyle\tabskip\z@{##}$\hfil
  \tabskip\@centering&\llap{##}\tabskip\z@\cr}
\def\endsneqnarray{\@@sneqncr\egroup $$\global\@ignoretrue}
\def\@@sneqncr{\let\@tempa\relax
   \ifcase\@eqcnt \def\@tempa{& & &}\or \def\@tempa{& &}
   \else \def\@tempa{&}\fi
     \@tempa \if@eqnsw\@eqnnum\stepcounter{subequation}\fi
     \global\@eqnswtrue\global\@eqcnt\z@\cr}

\makeatother

\begin{document}

\begin{titlepage}

\noindent
MPI--PhT/94--70 \hfill hep-th/9411001\\
CCNY--HEP--94/07\\
ITP--UH--15/94 \hfill October 1994\\

\vskip 0.6cm

\begin{center}

{\Large\bf  Four-Dimensional Twisted Group Lattices}\\

\vskip 1.0cm

{ Olaf Lechtenfeld$^{*}$
\, and \,
Stuart Samuel$^{\dagger\#}$}

\vskip 0.6cm

{\small
$^*${\it Institut f\"ur Theoretische Physik, Universit\"at Hannover}\\
{\it Appelstra\ss{}e 2, 30167 Hannover, Germany}\\
{E-mail: lechtenf@itp.uni-hannover.de}\\
[0.4cm]

$^\dagger$
{\it Max-Planck-Institut f\"ur Physik, Werner-Heisenberg-Institut}\\
{\it F\"ohringer Ring 6, 80805 Munich, Germany}\\
{E-mail: samuel@scisun.sci.ccny.cuny.edu}\\
[0.5cm]
}
\normalsize
\vskip 1cm
\textwidth 6truein
{\bf Abstract}
\end{center}
\begin{quote}
\hspace{\parindent}
{}\ \ \ Four-dimensional twisted group lattices
are used as models for space-time structure.
Compared to other attempts at space-time deformation,
they have two main advantages:
They have a physical interpretation
and there is no difficulty in putting field theories
on these structures.
We present and discuss ordinary and gauge theories
on twisted group lattices.
We solve the free field theory case
by finding all the irreducible representations.
The non-abelian gauge theory on the two-dimensional
twisted group lattice is also solved.
On twisted group lattices,
continuous space-time translational and rotational
symmetries are replaced by discrete counterparts.
We discuss these symmetries in detail.
Four-dimensional twisted group lattices
can also be used as models
for non-trivial discrete compactifactions
of certain ten-dimensional spaces.
\end{quote}

\vfill

\textwidth 6.5truein
\hrule width 5.cm

{\small
\noindent $^\#$
Permanent address: Department of Physics,
City College of New York,\\
\hspace*{0.4cm}New York, NY 10031, USA\\
}

\eject
\end{titlepage}

\secteqno

\section{Introduction}
\label{s:i}

\hspace{\parindent}
Recently, there has been interest in attempts
to deform the structure of space-time.
Normally, space-time coordinates
$x_1 , x_2 , \dots , x_d$ commute:
$x_j x_i = x_i x_j$.
One can $q$-deform this algebra
by assuming a non-commutative structure.
For example,
the quantum hyperspaces
\ct{manin88a,wz90a}
are defined by the algebra $x_j x_i = q_{ij} x_i x_j$,
for $i<j$,
where the $ q_{ij} $ are complex numbers.
In the $ q_{ij} \to 1$ limit,
one recovers ordinary space-time.%
{\footnote{One can consider more general relations,
such as
$ x_j x_i = {R_{ji}}^{kl} x_k x_l$,
where ${R_{ji}}^{kl}$ satisfy the Yang-Baxter equation.
Such spaces are even more difficult to interpret.}}

It is not completely clear
what a quantum hyperspace physically represents.
It appears to possess some kind of lattice structure.
Because of this lack of understanding,
field theories cannot yet be defined on quantum hyperspaces.
For this reason, contact with standard physics
is not yet possible.

In this work,
we shall consider a particular deformation
of space-time structure
which avoids the disadvantages of quantum hyperspaces.
We call these spaces twisted group lattices.
They are closely related to quantum hyperspaces
but have several physical interpretations.
Furthermore, one can define standard field theories
on twisted group lattices.
Twisted group lattices are, in some sense,
the natural extension of quantum hyperspaces.
In this work,
we focus on the four-dimensional twisted group lattice
and field-theoretic aspects.

The twisted group lattice is a lattice.
In general,
a lattice is determined by specifying sites and bonds.
By definition,
two sites are nearest neighbors
if there is a bond between them.

A group lattice
\ct{samuel90a,samuel91a}
is built from a discrete group $G$.
One declares a one-to-one correspondence
between the elements of $G$ and the sites of the lattice.
Hence the number of sites in the lattice is
the order of the group $ o ( G ) $, i.\ e.,
the total number of elements of the group.
By convention,
the origin corresponds to the identity element $e$ of $G$.
The nearest neighbor sites are determined
by a subset $\NN $ of elements of $G$,
satisfying the property that
if $h \in \NN $ then $h^{-1} \in \NN $.
Usually $\NN $ is not chosen to be a subgroup of $G$.
A site $g'$ is
a nearest neighbor site of $g$ if $g'g^{-1} \in \NN $.
The nearest neighbor sites of the origin $e$ are
the elements in $\NN $.
In general, the nearest neighbors of g are
$hg$ for $h \in \NN $.
Since the bonds ${\cal B}$ of the lattice are
lines between two nearest neighbor sites:
$
  {\cal B} =
  \{ [ hg , g ] \ , \hbox{ such that } g \in G, h \in \NN  \}
$.
Hence, a group lattice is
not only determined by $G$
but also by $\NN $.
For more details on group lattices,
see Refs.\ct{samuel90a}-\ct{samuel92a}.

For complicated $\NN $ and $G$,
the corresponding group lattice can appear quite complex.
However, a group lattice has more structure
than a random lattice.
The structure is governed by $G$.
Group lattices are homogeneous
in the sense that the lattice appears the same
when viewed from any site.

One can consider the free propagation of bosons or fermions
on a group lattice.
Such particles are allowed to hop from site to site.
Hopping parameters control the ease or difficulty
of passing over a bond.
One reason that group lattices are interesting is
that the corresponding statistical mechanics system
can be solved exactly
if the irreducible representations of $G$ are known.
The solution is given
in Ref.\ct{samuel90a}.
The solution method uses the group analog of the Fourier transform.
Furthermore,
if interactions are included
then a perturbative expansion of the theory is possible
because propagators exist and can be computed.
Thus, it is straightforward to put field theories
on group lattices and perform calculations.

Because the concept of a group lattice is general,
there should be many applications.
Indeed, the carbon atoms of the buckyball $C_{60}$ sit
at the sites of a group lattice based on the group $A_5$,
the alternating group on $5$ elements.
Using this result and some approximations,
one can compute the electronic structure of $C_{60}$.
The agreement of the group-lattice results
with experiments is good
\ct{samuel93a,samuel93b}.

The regular periodic hyper-cubic lattices in $d$-dimensions
of size $L_1 \times L_2 \times \dots \times L_d $
correspond to a group lattice based on an abelian group.
The group is
$G = Z_{L_1} \times Z_{L_2} \times \dots \times Z_{L_d}$.
Let $x_i$ be the generator of the $i$th abelian factor,
$Z_{L_i}$,
where $x_i^{L_i} = e$.
Let $\NN $  consist of the $x_i$ and their inverses, i.\ e.,
$
 \NN = \{ {
   x_1, x_1^{-1}, x_2, x_2^{-1}, \dots , x_d, x_d^{-1}
         } \}
$.
Each $g \in G$ is of the form
$g = x_1^{n_i}x_2^{n_2} \dots x_d^{n_d}$
with $0 \le n_i \le L_{i} - 1$.
Associate the point $(n_1, n_2, \dots , n_d)$
in the $d$-dimensional hyper-cubic lattice with this $g$.
Since the nearest neighbors to
$x_1^{n_i} x_2^{n_2} \dots x_d^{n_d}$
are obtained by multiplying by an $x_i$ or an $x_i^{-1}$,
the nearest neighbors of $(n_1, n_2, \dots , n_d)$ are
$(n_1, n_2, \dots , n_i \pm 1 , \dots , n_d) $.
Clearly, one obtains
the hyper-cubic periodic $d$-dimensional lattice.
The infinite hyper-cubic lattice is achieved
by replacing the $Z_{L_i}$ by the group of integers $Z$.
This can be achieved by taking the ${L_i} \to \infty$
and can be regarded as a thermodynamic limit.

The generator $x_i$ can be thought of
as taking a step in the $i$th direction.
Likewise $x_i^{-1}$ is a step in the negative $i$-direction.
The element
$g = x_1^{n_i} x_2^{n_2} \dots x_d^{n_d}$
is obtained by starting at the origin and
taking $n_1$ steps in the $1$-direction,
followed by $n_2$ steps in the $2$-direction, $\dots$,
and finally, $x_d$ steps in the $d$-direction.
Because the group is abelian,
it doesn't matter in which order one proceeds.

The $d$-dimensional twisted group lattice is based
on the ordinary $d$-dimensional hyper-cubic lattice.
One uses the above notion that $x_i$ and $x_i^{-1}$
correspond to steps in the plus or minus $i$-direction.
However, paths which go around closed loops
do not necessarily return to the origin $e$.
Instead one arrives at new group elements.
This introduces local twisting into the system.
Let $N_{ij}$, for $i<j$,
be a set of fixed positive integers.
When going around an elementary plaquette
in the $i$-$j$ plane,
one arrives at a new element $z_{ij}$
which satisfies $z_{ij}^{N_{ij}} = e$.
In other words,
$x_j^{-1} x_i^{-1} x_j x_i = z_{ij}$, for $i<j$.%
{\footnote{Our
convention is that the elements
associated with movements are multiplied
from right to left.
For example, $x_j x_i $ represents
first a movement in the $i$-direction
and then a movement in the $j$-direction.}}
The $z_{ij}$ commute with all elements of the group.
In moving around an elementary plaquette
in the $i$-$j$ plane,
one must go around it $N_{ij}$ times to return to the origin.
A general path returns to the origin
if and only if the region
projected onto each of the $i$-$j$ planes
has an area which is $0 \ ( \bmod \, {N_{ij}} )$.
The precise definition of $G$ is
$$
  G\ =\ \Bigl\{ { x_1^{n_1} x_2^{n_2} \ldots x_d^{n_d}  }
  \prod_{i<j} {z_{ij}^{n_{ij}}}
    \ , {\hbox { such that }} n_i = 0, 1, \ldots  ,L_i - 1 ,
$$
$$
   \qquad n_{ij}  =0,1,\ldots ,N_{ij}-1 ,
    \ x_j x_i x_j^{-1} x_i^{-1} =
       z_{ij} \hbox{ for } i<j ,
    \ x_k z_{ij} = z_{ij} x_k ,
$$
\be
   \qquad\qquad {z_{ij}z_{kl} = z_{kl}z_{ij} ,
    \ x_i^{L_i} =
    z_{ij}^{N_{ij}} = e ,
    \ N_{ij} \hbox{ divides } L_i \hbox{ and } L_j
   \hbox{ for all } i \hbox{ and } j } \Bigr\}
\quad .
\label{1.1}
\ee

Part of the interest in these twisted group lattices lies
in the connection with quantum hyperspaces
\ct{manin88a,wz90a}.
Indeed, if the $z_{ij}$ are represented by complex numbers,
one obtains the quantum hyperspaces,
of which the $d=2$ case is the quantum plane
\ct{manin88a}.
Different values of the $\{ N_{ij} \}$
and the $\{ L_i \}$
produce different twisted lattices.
The irreducible representations
for the general
two- and three-dimensional twisted group lattices
were found
in Refs.\ct{samuel91b,samuel92a}.
Hence, the propagation of free particles on those group lattices
has been exactly solved.
Propagation on twisted lattices is very different
from regular lattices because closed paths are obtained
only if the vanishing $\ (\bmod \, {N_{ij}})$ area constraints
are satisfied.

In this paper,
we solve the four-dimensional twisted group lattice.
Some motivation for studying this case
is that these lattices might be useful
in connection with four-dimensional physics.
For example,
one can consider field theories on group lattices.
An open question is whether such field theories play
a role in particle physics,
perhaps approximately or in a limit.
Other motivation comes
from quantum groups and quantum hyperspaces.
Because of the close similarity between
the twisted group lattices and quantum hyperspaces,
one might be able to gain insight into the latter.

The solution method requires finding
the irreducible representations
of the four-dimensional twisted group lattices.
Some elementary number theory plays a role.
In Sect.\rf{s:pf},
we perform a prime factorization of the system.
This allows one to solve the general problem
by analyzing smaller subsectors.
In Sect.\rf{s:cir},
we obtain all the irreducible representations.
This allows us to solve the free theory case.
Since we are interested in possible physical applications,
we consider field theories on generic group lattices
in Sect.\rf{s:fta}.
In particular, non-abelian gauge theories can be defined
using methods similar to those of Wilson
\ct{wilson}.
The gauge theory on the two-dimensional twisted lattice
can be solved exactly.
The solution is presented
in Sect.\rf{ss:lgtgl}.
Another topic concerns the gamma matrix structure
which arises for the $N_{ij}=2$ case in a particular subsector.
We find that Kogut-Susskind staggered fermions naturally arise.
In Sect.\rf{ss:stds},
the space-time discrete symmetries
of the twisted group lattices are analyzed.
At this early stage,
it is unclear whether group lattices
can play a role as a possible space-time structure.
In the Conclusion,
we discuss the relation to compactified models
and the possible implications for low-energy physics.
An Appendix presents the proof that
the representations obtained
in Sect.\rf{s:cir} are irreducible and complete.

We follow the notation
in Refs.\ct{samuel90a,samuel91b,samuel92a}.
A good introduction to the elementary number theory
which we use is Ref.\ct{leveque}.
For integers $a$ and $b$, $a | b$ indicates
that $a$ divides $b$, $\gcd (a, b)$ is
the greatest common division of $a$ and $b$, i.\ e.,
the biggest positive integer, $c$, such that $c|a$ and $c | b$,
and $\lcm (a, b) $ is the least common multiple of $a$ and $b$,
that is, the smallest integer, $c$, such that $a | c$ and $b | c$.
We work in Euclidean space.
Minkowski space can be obtained, in principle,
by Wick rotation:
For scalar field theories,
one changes the sign of the kinetic energy terms
involving time derivatives
and one includes a factor of $i$ in the action.

Other notation is as follows.
A subscript in parenthesis on a $P$ or $Q$ matrix indicates
the size of the representation
and signifies that commuting
$P_{\left( N \right)}$ past $Q_{\left( N \right)}$
produces the phase
$\exp \left( {{{2\pi i} \over N}} \right)$,
as in Eq.\ \bref{2.1}.
The integer $\ell$ labels the prime-factor sectors.
When it appears as a superscript,
we enclose it in parenthesis
to avoid confusion with a power.
The primes appearing in the prime factorization
are denoted by $p_\ell$.
The number of distinct primes is $L$.
The variables $r, s, t, u$
with subscripts or superscripts
denote
integer powers.
The symbol $r$ without a subscript
indicates an irreducible representation.
The variables $i,j, k, l$ take on the values $1, 2, 3,$ and $4$
and denote space-time directions.
In the combination $\pi i$,
$i$ represents $\sqrt{-1}$ and not a space-time direction.

\section{Prime Factorization}
\label{s:pf}

\hspace{\parindent}
In this section,
we introduce an idea
which allows one to simplify the construction
of the irreducible representations
of the group associated with the $d$-dimensional twisted lattice.
The point is to factorize the problem
into a series of smaller parts.

Fundamental to the construction
of twisted-group-lattice representations are
the $N \times N$ matrices
$P_{\left( N \right)}$ and $Q_{\left( N \right)}$.
They satisfy the commutation relations
\be
  P_{\left( N \right)} Q_{\left( N \right)}\ =\
  Q_{\left( N \right)} P_{\left( N \right)}
    \exp \left( {{{2\pi i} \over N}} \right)
\quad .
\label{2.1}
\ee
Explicit matrix forms for
$P_{\left( N \right)}$ and $Q_{\left( N \right)}$
are
\ct{thooft79a}
$$
  P_{(N)}\ =\
    \left( \matrix{
   0  &1  &0  &0  &\ldots   &0  &0  \cr
   0  &0  &1  &0  &\ldots   &0  &0  \cr
   0  &0  &0  &1  &\ldots   &0  &0  \cr
   \vdots  &\vdots  &\vdots  &\vdots  &\ddots   &\vdots  &\vdots  \cr
   0  &0  &0  &0  &\ldots   &0  &1  \cr
   1  &0  &0  &0  &\ldots   &0  &0  \cr
    } \right)
\quad ,
$$
\be
  Q_{(N)}\ =\
    \left( \matrix{
   1  &0  &0  &0  &\dots   &0  \cr
   0  &\ex{2}  &0  &0  &\ldots   &0  \cr
   0  &0  &\ex{4}  &0  &\ldots   &0  \cr
   \vdots  &\vdots  &\vdots  &\vdots  &\ddots   &\vdots  \cr
   0  &0  &0  &0  &\ldots   &\ex{2(N-1)}  \cr
    } \right)
\quad .
\label{2.2}
\ee

Suppose
\be
  N = N_1 N_2
\quad ,
\label{2.3}
\ee
where $N_1$ and $N_2$ are relatively prime.
It is an elementary result
\ct{leveque}
of number theory
that there then exist integers $a_1$ and $a_2$
such that
\be
  a_1 N_1 + a_2 N_2 = 1
\quad .
\label{2.4}
\ee
One necessarily has
\be
    \gcd \left( {a_1,N_2} \right)
  = \gcd \left( {a_2,N_1} \right) = 1
\quad .
\label{2.5}
\ee
An algorithm
\ct{leveque}
exists
for obtaining $a_1$ and $a_2$,
given $N_1$ and $N_2$.
When  $N_1$ and $N_2$ are not large,
it is usually easy to find
$a_1$ and $a_2$
by inspection.
{}From Eqs.\ \bref{2.3} and \bref{2.4},
it follows that
\be
  {1 \over N} = {{a_2} \over {N_1}} + {{a_1} \over {N_2}}
\quad .
\label{2.6}
\ee
Using this result and Eq.\ \bref{2.1},
it is easily shown that
$P_{\left( N \right)}$ and $Q_{\left( N \right)}$
can be represented in tensor product form as
$$
  Q_{\left( {N_1} \right)} \times Q_{\left( {N_2} \right)}
\quad ,
$$
\be
  P_{\left( {N_1} \right)}^{a_2}
  \times P_{\left( {N_2} \right)}^{a_1}
\quad .
\label{2.7}
\ee

Continuing the process,
one factors $N$ over the primes
via
\be
  N = \prod_{\ell=1}^L {p_{\ell}^{s_{\ell}}}
\quad ,
\label{2.8}
\ee
where $p_1 , p_2 , \ldots , p_L$ are the $L$ distinct primes
appearing in $N$.
The power of $p_{\ell}$ in $N$ is $s_{\ell}$.
Again,
by repeating the construction in the previous paragraph,
there exist $b_{\ell}$,
$\ell = 1, 2, \ldots , L$,
satisfying
\be
  \gcd \left( {b_{\ell},p_{\ell}^{s_{\ell}}} \right) = 1
\quad ,
\label{2.9}
\ee
and such that
\be
  \sum\limits_{\ell=1}^L
  {{{b_{\ell}} \over {p_{\ell}^{s_{\ell}}}}} = {1 \over N}
\quad .
\label{2.10}
\ee
Then,
$P_{\left( N \right)}$ and $Q_{\left( N \right)}$
can be represented as a tensor product of terms
over the prime factors
via
$$
  Q_{\left( {p_1^{s_1}} \right)} \times
  Q_{\left( {p_2^{s_2}} \right)} \times \ldots \times
  Q_{\left( {p_L^{s_L}} \right)}
\quad ,
$$
\be
  P_{\left( {p_1^{s_1}} \right)}^{b_1} \times
  P_{\left( {p_2^{s_2}} \right)}^{b_2} \times \ldots \times
  P_{\left( {p_L^{s_L}} \right)}^{b_L}
\quad .
\label{2.11}
\ee
As side remark, we note that,
since the representation
of the algebra
in Eq.\ \bref{2.1}
is unique,
the matrices
in Eqs.\ \bref{2.2} and \bref{2.11}
are necessarily equivalent via a conjugation.

The advantage of using prime factorization
is that the problem of constructing
the irreducible representations
for $d$-dimensional twisted group lattice
also factorizes.
In other words,
it suffices to find irreducible representations
for a single prime-factor sector.

\vfill\eject

\section{Construction
of the Irreducible Representations}
\label{s:cir}

\hspace{\parindent}
The goal of this section is to find
the complete set of irreducible representations
for the group associated with the 4-d twisted lattice.
With this knowledge,
the problem of free propagation
of particles on the lattice can be solved
and a perturbative expansion of an interacting theory
can be performed.

The parameters we use to characterize a representation
are internal momenta $k_{ij}$ and space-time momenta $k_{i}$.
There is an irreducible representation $r$
for each value of the ``momenta''
$
   k_{ij} = 0,1,\ldots , N_{ij}-1
$
associated with the $z_{ij}$ group elements.
Here, the indices $ij$ take on the values
$ ij=12, 13, 14, 23, 24,$ and $34 $.
When curly brackets $\{ \ \}$ appear
around a variable with $ij$ indices,
it means that $ij$ range over these six values.
Let $d_r$ denote
the dimension of the representation $r$.
We choose the representation so that
the matrices associated with $z_{ij}$ are diagonal
and proportional to the identity matrix $I_{(d_r)}$, i.\ e.,
\be
  z_{ij} \rightarrow I_{(d_r)}
   \exp \left( { 2 \pi i { {k_{ij} } \over { N_{ij} } } } \right)
\quad .
\label{3.1}
\ee
This is possible because the $z_{ij}$ mutually commute.
Note that the matrix representation of $z_{ij}$
in Eq.\ \bref{3.1}
raised to the power $ N_{ij}$ yields the identity matrix,
as it should,
since $z_{ij}^{ N_{ij}}=1$.

Representations are also determined by the ``momenta''
$
   k_i = 0,1, \ldots ,L_{i}-1
$
associated with the elements $x_i$,
where $i=1,2,3$ or $4$.
However, some sets of $\{ k_1 ,  k_2 , k_3 , k_4 \}$
lead to equivalent representations.
Such representations are related
by an element of a group $E$.
Below we specify $E$ in detail.
The matrix representations of $x_j$
are of the form
\ct{samuel92a}
\be
  x_j \rightarrow \Gamma_j
   \exp \left( { 2 \pi i { {k_{j} } \over { L_{j} } } } \right)
\quad ,
\label{3.2}
\ee
where the $\Gamma_i$ are
$d_r \times d_r$ matrices
satisfying
\be
  \Gamma_j \Gamma_i\ =\ \Gamma_i \Gamma_j
    \exp \left( {2\pi i{{k_{ij}} \over {N_{ij}}}} \right)
\ , \quad \quad
{\hbox {for}} \ 1 \le i < j \le 4
\quad ,
\label{3.3}
\ee
and
\be
  \Gamma_i^{L_i} = 1
\ , \quad \quad
{\hbox {for}} \ i = 1, 2, 3, {\hbox { or }} 4
\quad .
\label{3.4}
\ee

For the rest of this section,
we assume that a specific
set of $\{ k_{ij} \}$ is given.
Write
\be
  {{k_{ij}}  \over {N_{ij}}} =
  {{k_{ij}'} \over {N_{ij}'}}
\quad ,
\label{3.5}
\ee
where the prime indicates that
the fraction has been reduced, i.\ e.,
common factors have been cancelled so that
\be
  \gcd \left( {k_{ij}',N_{ij}'} \right) = 1
\quad .
\label{3.6}
\ee
When $k_{ij} = 0$,
we set $k_{ij}' = 0$ and $N_{ij}' = 1$.
Note that it is only the combination
$
  {{k_{ij}}  / {N_{ij}}} =
  {{k_{ij}'} / {N_{ij}'}}
$ which enters
Eqs.\ \bref{3.1} and \bref{3.3}
so that the primed variables
are the quantities of interest.

Define
\be
  N' \equiv \lcm \left( { \{ { N_{ij}' } \} } \right)
  = \lcm \left( {N_{12}',N_{13}',N_{14}',
                 N_{23}',N_{24}',N_{34}'} \right)
\quad .
\label{3.7}
\ee
Perform a prime factorization of $N'$
\be
  N' = \prod_{\ell=1}^L {p_{\ell}^{s_{\ell}}}
\ , \quad \quad
{\hbox {where}} \ s_{\ell} \ge 1
\quad ,
\label{3.8}
\ee
and of the $N_{ij}'$
\be
  N_{ij}' =
     \prod_{\ell=1}^L
        {p_{\ell}^{t_{ij}^{\left( {\ell} \right)}}}
\ , \quad \quad
{\hbox {where}} \ t_{ij}^{\left( {\ell} \right)} \ge 0
\quad .
\label{3.9}
\ee
The integer power
$
  t_{ij}^{\left( {\ell} \right)}
$
takes on the value zero
when the prime $p_{\ell}$
does not appear in $N_{ij}'$.
The exponent $s_{\ell}$ is
the maximum value
of $t_{ij}^{\left( {\ell} \right)}$
as $ij$ ranges over $1 \le i < j \le 4$:
\be
  s_{\ell} = \mathop{\max }\limits_{\left\{ {ij} \right\}}
   \left\{ {t_{ij}^{\left( {\ell} \right)}} \right\}
\quad .
\label{3.10}
\ee
Next define
$
  b_{ij}^{\left( {\ell} \right)}
$
by
\be
  \sum\limits_{\ell=1}^L
  {{{b_{ij}^{\left( {\ell} \right)}} \over {p_{\ell}^{s_{\ell}}}}} =
   {{k_{ij}'} \over {N_{ij}'}}
\quad .
\label{3.11}
\ee
Note that $b_{ij}^{\left( {\ell} \right)}$ may divide
$p_{\ell}^{s_{\ell}}$
so that
$
  \gcd \left( {b_{ij}^{\left( {\ell} \right)} ,
              p_{\ell}^{s_{\ell}}}
     \right)
$
is not necessarily $1$.
However, if
$
  t_{ij}^{\left( {\ell} \right)} =  s_{\ell}
$
for specific $ij$
then
$
  \gcd \left( {
             b_{ij}^{\left( {\ell} \right)} ,
             p_{\ell}^{s_{\ell}}
               }  \right) = 1
$.
Thus, Eq.\ \bref{3.10}
implies that at least one
$b_{ij}^{\left( {\ell} \right)}$
satisfies
$
  \gcd \left( {
             b_{ij}^{\left( {\ell} \right)} ,
             p_{\ell}^{s_{\ell}}
               }  \right) = 1
$.

The strategy is to solve the problem
for a prime-factor sector $\ell$.
Write $\Gamma_i$ as a tensor product via
\be
  \Gamma_i\ =\            \Gamma_i^{\left( 1 \right)}
                 \times \Gamma_i^{\left( 2 \right)}
   \times \ldots \times \Gamma_i^{\left( L \right)}
\quad .
\label{3.12}
\ee
If
\be
  \Gamma_j^{\left( {\ell} \right)}
  \Gamma_i^{\left( {\ell} \right)}\ =\
  \Gamma_i^{\left( {\ell} \right)}
  \Gamma_j^{\left( {\ell} \right)}
  \exp \left( {2\pi i
  {{b_{ij}^{\left( {\ell} \right)}} \over {p_{\ell}^{s_{\ell}}}}}
       \right)
\quad ,
\label{3.13}
\ee
and
\be
  \left( {\Gamma_i^{\left( {\ell} \right)}} \right)^{L_i} = 1
\quad ,
\label{3.14}
\ee
then Eqs.\ \bref{3.3} and \bref{3.4}
are satisfied due to
Eqs.\ \bref{3.5} and \bref{3.11} -- \bref{3.14}.
Define
\be
  p_{\ell}^{r_{\ell}}\
  \equiv\ \gcd
  \left( {b_{12}^{\left( {\ell} \right)}
          b_{34}^{\left( {\ell} \right)} -
          b_{13}^{\left( {\ell} \right)}
          b_{24}^{\left( {\ell} \right)} +
          b_{14}^{\left( {\ell} \right)}
          b_{23}^{\left( {\ell} \right)}\ ,\
     p_{\ell}^{s_{\ell}}} \right)
\quad ,
\label{3.15}
\ee
and let
\be
  u_{\ell} \equiv s_{\ell} - r_{\ell}
\  ,
\quad \quad
  d_r^{\left( {\ell} \right)}
   \equiv p_{\ell}^{2s_{\ell} - r_{\ell}} =
           p_{\ell}^{s_{\ell}} p_{\ell}^{u_{\ell}}
\quad .
\label{3.16}
\ee
The dimension of the matrix
$\Gamma_i^{\left( {\ell} \right)}$ is
\be
  \dim \Gamma_i^{\left( {\ell} \right)} =
    d_r^{\left( {\ell} \right)}
\quad .
\label{3.17}
\ee
The quantity
$
  b_{12}^{\left( {\ell} \right)}
  b_{34}^{\left( {\ell} \right)} -
  b_{13}^{\left( {\ell} \right)}
  b_{24}^{\left( {\ell} \right)} +
  b_{14}^{\left( {\ell} \right)}
  b_{23}^{\left( {\ell} \right)}
$
is related to the ``obstruction''
of the $\ell$-sector matrices
to be of minimum size.
If
$$
  b_{12}^{\left( {\ell} \right)}
  b_{34}^{\left( {\ell} \right)} -
  b_{13}^{\left( {\ell} \right)}
  b_{24}^{\left( {\ell} \right)} +
  b_{14}^{\left( {\ell} \right)}
  b_{23}^{\left( {\ell} \right)}\ =\ 0
   \pmod { p_{\ell}^{s_{\ell}} }
\quad ,
$$
then $ u_{\ell} = 0 $
and the minimum dimension
$p_{\ell}^{s_{\ell}}$ occurs.
As will be evident below,
the matrices $\Gamma_i$ are tensor products
of two matrices of size
$ N_{\ell}' \times  N_{\ell}'$ and
$M_{\ell} \times M_{\ell}$, where
\be
  N_{\ell}' \equiv
  p_{\ell}^{s_{\ell}}
\quad , \quad \quad
  M_{\ell} \equiv p_{\ell}^{u_{\ell}}
\quad .
\label{3.19}
\ee
The total dimension is
the product of the dimensions for each prime sector:
\be
  d_r = \prod_{\ell=1}^L {d_r^{\left( {\ell} \right)}} =
        \prod_{\ell=1}^L
           p_{\ell}^{s_{\ell}} p_{\ell}^{u_{\ell}}
\quad .
\label{3.20}
\ee

The construction of the representation
depends on which
$b_{ij}^{\left( {\ell} \right)}$
has
$
  \gcd \left( {
             b_{ij}^{\left( {\ell} \right)} ,
             p_{\ell}^{s_{\ell}}
               }  \right) = 1
$.
There are six cases

(1)
$
  \gcd \left( {b_{12}^{\left( {\ell} \right)},p_{\ell}} \right) = 1
$,

(2)
$
  \gcd     \left( {b_{13}^{\left( {\ell} \right)},p_{\ell}} \right) = 1
$,
  but
$
 \gcd \left( {b_{12}^{\left( {\ell} \right)},p_{\ell}} \right) > 1
$,

(3)
$
  \gcd     \left( {b_{14}^{\left( {\ell} \right)},p_{\ell}} \right) = 1
$,
  but
$
  \gcd \left( {b_{12}^{\left( {\ell} \right)},p_{\ell}} \right) > 1
$
  and
$
 \gcd \left( {b_{13}^{\left( {\ell} \right)},p_{\ell}} \right) > 1
$,

(4)
$
  \gcd     \left( {b_{23}^{\left( {\ell} \right)},p_{\ell}} \right) = 1
$,
  but
$     \gcd \left( {b_{1j}^{\left( {\ell} \right)},p_{\ell}} \right) > 1
$,
for $j=2, 3,$ and $4$,

(5)
$
  \gcd     \left( {b_{24}^{\left( {\ell} \right)},p_{\ell}} \right) = 1
$,
  but
$
\gcd \left( {b_{1j}^{\left( {\ell} \right)},p_{\ell}} \right) > 1
$,
for $j=2, 3, 4$,
and
$
      \gcd \left( {b_{23}^{\left( {\ell} \right)},p_{\ell}} \right) > 1
$,

(6)
$
  \gcd \left( {b_{34}^{\left( {\ell} \right)},p_{\ell}} \right) = 1
$,
 but
$
  \gcd \left( {b_{ij}^{\left( {\ell} \right)},p_{\ell}} \right) > 1
$
for all $i<j$ except $ij=34$.

We treat case (1) in detail.
The other five cases are similar
and can be obtained from case (1)
by permuting space-time indices.
For case (1),
\be
  \gcd
  \left( {
     b_{12}^{\left( {\ell} \right)} ,
     p_{\ell}^{s_{\ell}}
         } \right) = 1
\quad .
\label{3.21}
\ee
Let
$$
  \Gamma_1^{\left( {\ell} \right)} =
    Q_{\left( {N_{\ell}'} \right)}
   \times I_{\left( {M_{\ell}} \right)}
\quad ,
$$
\be
  \Gamma_2^{\left( {\ell} \right)} =
   P_{\left( {N_{\ell}'} \right)}^{b_{12}^{\left( {\ell} \right)}}
   \times I_{\left( {M_{\ell}} \right)}
\quad .
\label{3.22}
\ee
Equation \bref{3.21} implies that there exist integers
$a_3^{\left( {\ell} \right)}$ and
$a_4^{\left( {\ell} \right)}$
such that
$$
   -b_{12}^{\left( {\ell} \right)}
       a_3^{\left( {\ell} \right)} =
    b_{23}^{\left( {\ell} \right)}
   \pmod { p_{\ell}^{s_{\ell}} }
\quad ,
$$
\be
   -b_{12}^{\left( {\ell} \right)}
       a_4^{\left( {\ell} \right)} =
    b_{24}^{\left( {\ell} \right)}
   \pmod { p_{\ell}^{s_{\ell}} }
\quad .
\label{3.23}
\ee
Equations \bref{3.15}, \bref{3.16} and \bref{3.23}
imply
\be
 {{b_{34}^{\left( {\ell} \right)} +
      a_4^{\left( {\ell} \right)}
   b_{13}^{\left( {\ell} \right)} -
      a_3^{\left( {\ell} \right)}
   b_{14}^{\left( {\ell} \right)}} \over {p_{\ell}^{s_{\ell}}}}\ =\
   {{b_4^{\prime \left( {\ell} \right)}}
       \over {p_{\ell}^{s_{\ell} - r_{\ell}}}}\ =\
   {{b_4^{\prime \left( {\ell} \right)}} \over {p_{\ell}^{u_{\ell}}}}
\quad ,
\label{3.24}
\ee
where
\be
  \gcd \left( {b_4^{\prime \left( {\ell} \right)},
     p_{\ell}^{u_{\ell}}} \right) = 1
\quad .
\label{3.25}
\ee
The two remaining matrices turn out to be given by
$$
  \Gamma_3^{\left( {\ell} \right)}\ =\
     Q_{\left( {N_{\ell}'} \right)}^{a_3^{\left( {\ell} \right)}}
     P_{\left( {N_{\ell}'} \right)}^{b_{13}^{\left( {\ell} \right)}}
    \times Q_{\left( {M_{\ell}} \right)}\
    \exp \left[ {-i\pi \left( {L_3{-}1} \right)
       {{a_3^{\left( {\ell} \right)}
      b_{13}^{\left( {\ell} \right)}} \over {N_{\ell}'}}} \right]
\quad ,
$$
\be
  \Gamma_4^{\left( {\ell} \right)}\ =\
    Q_{\left( {N_{\ell}'} \right)}^{a_4^{\left( {\ell} \right)}}
    P_{\left( {N_{\ell}'} \right)}^{b_{14}^{\left( {\ell} \right)}}
    \times P_{\left( {M_{\ell}} \right)
         }^{b_4^{\prime \left( {\ell} \right)}}\
    \exp \left[ {-i\pi \left( {L_4{-}1} \right)
    {{a_4^{\left( {\ell} \right)}
    b_{14}^{\left( {\ell} \right)}} \over {N_{\ell}'}}} \right]
\quad ,
\label{3.26}
\ee
where the phase factors ensure that
$
  \left( {\Gamma_i^{\left( {\ell} \right)}} \right)^{L_i} = 1
$.

It is straightforward to verify
that the $\Gamma_i^{\left( {\ell} \right)}$
in Eqs.\ \bref{3.22} and \bref{3.26}
satisfy Eq.\ \bref{3.13}:
For $i=1$ and $ j=2,3,4 $,
one immediately obtains
\be
  \Gamma_j^{\left( {\ell} \right)}
  \Gamma_1^{\left( {\ell} \right)}\ =\
  \Gamma_1^{\left( {\ell} \right)}
  \Gamma_j^{\left( {\ell} \right)}
   \exp \left( {2\pi i{{b_{1j}^{\left( {\ell} \right)}} \over
        {p_{\ell}^{s_{\ell}}}}} \right)
\quad .
\label{3.27}
\ee
For $i=2$ and $ j=3,4$,
\be
  \Gamma_j^{\left( {\ell} \right)}
  \Gamma_2^{\left( {\ell} \right)}\ =\
  \Gamma_2^{\left( {\ell} \right)}
  \Gamma_j^{\left( {\ell} \right)}
  \exp \left( {-2\pi i{{b_{12}^{\left( {\ell} \right)}
     a_j^{\left( {\ell} \right)}}
        \over {p_{\ell}^{s_{\ell}}}}} \right)\ =\
  \Gamma_2^{\left( {\ell} \right)}
  \Gamma_j^{\left( {\ell} \right)}
  \exp \left( {2\pi i{{b_{2j}^{\left( {\ell} \right)}} \over
     {p_{\ell}^{s_{\ell}}}}} \right)
\label{3.28}
\ee
follows
from Eq.\ \bref{3.23}.
Finally,
$$
  \Gamma_4^{\left( {\ell} \right)}
  \Gamma_3^{\left( {\ell} \right)}\ =\
  \Gamma_3^{\left( {\ell} \right)}
  \Gamma_4^{\left( {\ell} \right)}
   \exp \left( {2\pi i\left( {{{a_3^{\left( {\ell} \right)}
    b_{14}^{\left( {\ell} \right)}-a_4^{\left( {\ell} \right)}
    b_{13}^{\left( {\ell} \right)}} \over {p_{\ell}^{s_{\ell}}}} +
    {{b_4^{\prime \left( {\ell} \right)}} \over
      {p_{\ell}^{u_{\ell}}}}} \right)} \right)
$$
\be
 =\ \Gamma_3^{\left( {\ell} \right)}
   \Gamma_4^{\left( {\ell} \right)}
    \exp \left( {2\pi i{{b_{34}^{\left( {\ell} \right)}} \over
       {p_{\ell}^{s_{\ell}}}}} \right)
\quad ,
\label{3.29}
\ee
since
\be
 {{b_{34}^{\left( {\ell} \right)}} \over {p_{\ell}^{s_{\ell}}}}\ =\
   {{b_4^{\prime \left( {\ell} \right)}}
       \over {p_{\ell}^{u_{\ell}}}} +
    {{a_3^{\left( {\ell} \right)} b_{14}^{\left( {\ell} \right)} -
      a_4^{\left( {\ell} \right)}
   b_{13}^{\left( {\ell} \right)}} \over
               {p_{\ell}^{s_{\ell}}}}
\quad .
\label{3.30}
\ee
Equation \bref{3.30}
is a consequence
of Eq.\ \bref{3.24}.

All six cases lead to matrices of the form
\be
  \Gamma_i^{\left( {\ell} \right)}\ =\
    Q_{\left( {N_{\ell}'} \right)}^{a_i^{\left( {\ell} \right)}}
    P_{\left( {N_{\ell}'} \right)}^{b_i^{\left( {\ell} \right)}}
  \times
  Q_{\left( {M_{\ell}} \right)}^{a_i^{\prime \left( {\ell} \right)}}
  P_{\left( {M_{\ell}} \right)}^{b_i^{\prime \left( {\ell} \right)}}\
    \exp \left[ {-i\pi \left( {L_i{-}1} \right)
     \left( {{{a_i^{\left( {\ell} \right)}
      b_i^{\left( {\ell} \right)}} \over {N_{\ell}'}} +
    {{a_i^{\prime \left( {\ell} \right)}
      b_i^{\prime \left( {\ell} \right)}} \over
       {M_{\ell}}}} \right)} \right]
\quad .
\label{3.31}
\ee
For example, for case (1)
\begin{eqnarray}
  \matrix{
  a_1^{ \left( {\ell} \right)} = 1 \ ,\quad\hfill &
  b_1^{ \left( {\ell} \right)} = 0 \ ,\quad\hfill &
  a_1^{\prime \left( {\ell} \right)} = 0 \ ,\quad\hfill &
  b_1^{\prime \left( {\ell} \right)} = 0 \ ,\quad\hfill \cr
  a_2^{ \left( {\ell} \right)} = 0 \ ,\quad\hfill &
  b_2^{ \left( {\ell} \right)} = b_{12}^{\left({\ell}\right)}
\ ,\quad\hfill &
  a_2^{\prime \left( {\ell} \right)} = 0 \ ,\quad\hfill &
  b_2^{\prime \left( {\ell} \right)} = 0 \ ,\quad\hfill \cr
  a_3^{ \left( {\ell} \right)} = a_3^{\left({\ell}\right)}
\ ,\quad\hfill &
  b_3^{ \left( {\ell} \right)} = b_{13}^{\left({\ell}\right)}
\ ,\quad\hfill &
  a_3^{\prime \left( {\ell} \right)} = 1 \ ,\quad\hfill &
  b_3^{\prime \left( {\ell} \right)} = 0 \ ,\quad\hfill \cr
  a_4^{ \left( {\ell} \right)} = a_4^{ \left({\ell}\right)}
\ ,\quad\hfill &
  b_4^{ \left( {\ell} \right)} = b_{14}^{\left({\ell}\right)}
\ ,\quad\hfill &
  a_4^{\prime \left( {\ell} \right)} = 0 \ ,\quad\hfill &
  b_4^{\prime \left( {\ell} \right)} =
  b_4^{\prime \left( {\ell} \right)}
  \ . \quad\hfill \cr }
\nonumber \\
\quad
\label{3.32}
\end{eqnarray}
A fairly compact formula for the dimension $d_r$
can be obtained.
Abbreviating
\be
  n_{ij}' \equiv { {N'} \over  {N_{ij}'} }
\label{3.33}
\quad ,
\ee
one has
\be
  d_r = { {N^{\prime 2} } \over
  {\gcd \left( { n_{12}' k_{12}' n_{34}' k_{34}' -
                 n_{13}' k_{13}' n_{24}' k_{24}' +
                 n_{14}' k_{14}' n_{23}' k_{23}' \
  ,\ N'} \right)} }
\quad .
\label{3.34}
\ee

As remarked above,
some $\{ k_1 ,  k_2 , k_3 , k_4 \}$
lead to equivalent representations.
Such momentum sets are related by the action
of elements of a group $E$.
The group $E$ is generated by $4L$ elements.
The generators are
$$
 E_{P\times I}^{\left( \ell \right)}:
  {\hbox{ conjugation by }}
 P_{\left( {N_{\ell}'} \right)} \times
 I_{\left( {M_{\ell} } \right)}
  {\hbox{ on the }} \ell {\hbox{th factor}}
\quad ,
$$
$$
  E_{Q^{-1}\times I}^{\left( \ell \right)}:
   {\hbox{ conjugation by }}
  Q_{\left( {N_{\ell}'} \right)}^{-1} \times
  I_{\left( {M_{\ell} } \right)}
   {\hbox{ on the }} \ell {\hbox{th factor}}
\quad ,
$$
$$
  E_{I\times P}^{\left( \ell \right)}:
   {\hbox{ conjugation by }}
  I_{\left( {N_{\ell}'} \right)} \times
  P_{\left( {M_{\ell} } \right)}
   {\hbox{ on the }} \ell {\hbox{th factor}}
\quad ,
$$
\be
  E_{I\times Q^{-1}}^{\left( \ell \right)}:
   {\hbox{ conjugation by }}
  I_{\left( {N_{\ell}'} \right)} \times
  Q_{\left( {M_{\ell} } \right)}^{-1}
   {\hbox{ on the }} \ell {\hbox{th factor}}
\quad .
\label{3.35}
\ee
It is not difficult to see that the effect
of these generators is to
shift momenta by the following.
$$
  {\hbox {For}}\ E_{P\times I}^{\left( \ell \right)}:\qquad
  k_i\ \to\ k_i +
   {{a_i^{\left( {\ell} \right)}  L_i} \over {N_{\ell}'}}
   \pmod {L_i}
\quad ,
$$
$$
  {\hbox {for}}\ E_{Q^{-1}\times I}^{\left( \ell \right)}:\qquad
  k_i\ \to\ k_i +
   {{b_i^{\left( {\ell} \right)}  L_i} \over {N_{\ell}'}}
   \pmod {L_i}
\quad ,
$$
$$
  {\hbox {for}}\ E_{I\times P}^{\left( \ell \right)}:\qquad
  k_i\ \to\ k_i +
    {{a_i^{\prime \left( {\ell} \right)} L_i} \over {M_{\ell}}}
   \pmod {L_i}
\quad ,
$$
\be
  {\hbox {for}}\ E_{I\times Q^{-1}}^{\left( \ell \right)}:\qquad
  k_i\ \to\ k_i +
    {{b_i^{\prime \left( {\ell} \right)} L_i} \over {M_{\ell}}}
   \pmod {L_i}
\quad .
\label{3.36}
\ee
For fixed $\ell$,
$
  {p_{\ell}^{2s_{\ell}} p_{\ell}^{2u_{\ell}}}
  = \left( d_r^{\left( {\ell} \right)} \right)^2
$
identifications are made.
The identifications are independent
for different values of $\ell$.

Hence, the irreducible representations
are characterized by the six integer values of $k_{ij}$,
which range from $0$ to $N_{ij} - 1$,
and the values of the $k_i$ modulo $E$.
For fixed $\{ k_{ij} \}$,
the dimensions of the representations are the same, and
the total number of identifications of
the $\{ k_1 ,  k_2 ,  k_3 ,  k_4 \}$
under $E$ is equal to this dimension squared since
\be
  \prod_{\ell=1}^L
    {p_{\ell}^{2s_{\ell}} p_{\ell}^{2u_{\ell}}}
  = d_r^2
\quad .
\label{3.37}
\ee
Since each $k_i$ ranges from $0$ to $L_i - 1$,
there are
\be
 { {L_1 L_2 L_3 L_4} \over {d_r^2} }
\label{3.38}
\ee
independent values of $\{ k_1 ,  k_2 ,  k_3 ,  k_4 \}$
for specific $\{ k_{ij} \}$
($p_{\ell}$, $s_{\ell}$, $u_{\ell}$ and $d_r$
depend only on the $k_{ij}$).
It follows that
$$
  \sum\limits_{\left( r \right)} {d_r^2}\ =\
  \sum\limits_{k_{12}=0}^{_{N_{12}-1}} \ldots
  \sum\limits_{k_{34}=0}^{_{N_{34}-1}}
 \ \ \sum\limits_{\left\{ {k_1,k_2,k_3,k_4} \right\} \pmod E} {d_r^2}
$$
$$
  \qquad\qquad =\ \sum\limits_{k_{12}=0}^{_{N_{12}-1}} \ldots
  \sum\limits_{k_{34}=0}^{_{N_{34}-1}}
  \sum\limits_{k_1=0}^{_{L_1-1}}
  \ldots \sum\limits_{k_4=0}^{_{L_4-1}}
  {{L_1 L_2 L_3 L_4} \over {d_r^2}} d_r^2
$$
\be
  \phantom{.}\quad\qquad\qquad\qquad\qquad =\
  N_{12} N_{13} N_{14} N_{23} N_{24} N_{34} L_1 L_2 L_3 L_4\
  =\ o \left( G \right)
\quad ,
\label{3.39}
\ee
where $o \left( G \right) $ is the order of $G$, i.\ e.,
the number of elements in $G$.

\section{Field Theory Aspects}
\label{s:fta}

\hspace{\parindent}
In this section,
we consider field theories on group lattices.
We first treat the propagation of a free particle
on a four-dimensional twisted group lattice.
The problem can be solved exactly using the results
of Sect.\rf{s:cir}.
Secondly,
we show how gauge theories can be defined
on generic group lattices.
The case of the two-dimensional twisted group lattice
is solved exactly.
Next,
we consider anticommuting fields on the $\{ N_{ij} = 2 \}$
$d$-dimensional twisted group lattice.
It was noted
in Ref.\ct{samuel92a}
that Dirac-gamma-matrix structure naturally appears
in the $\{ k_{ij} = 1 \}$ sector.
We show that Kogut-Susskind staggered fermions,
and not some other type of lattice fermions, arise.
Finally,
we analyze the space-time discrete symmetries
of scalar field theories
on a $d$-dimensional twisted group lattice.

This section provides the foundation
for analyzing potentially interesting field theories
on four-dimensional twisted group lattices.
Standard-model-like particle theories
can be treated.
A gauge theory such as $SU(3) \times SU(2) \times U(1) $
can be defined on a four-dimensional twisted group lattice.
Matter fields can also be introduced and coupled to gauge fields.
A perturbative treatment of the system can be performed.
The first step in a perturbative expansion
is the solution of the free system.

\subsection{The Free Theory Solution}
\label{ss:fts}

\hspace{\parindent}
It is straightforward to put field theories
on a group lattice.
Local field theories involve the product
of fields at nearest neighbor sites
and at the same site.
For this reason,
it is convenient to append $e$ to $\NN$,
and define this set to be $\NN_e$:
\be
  \NN_e \equiv \NN \cup \left\{ e \right\}
\quad .
\label{4.1}
\ee

Consider a single charged scalar field, $\phi$.
A free action $S_0$ for $\phi$ is
\be
  S_0 = \sum\limits_{g\in G}
  \sum\limits_{h\in \NN_e}
  \lambda_h \phi^*\left( {hg} \right)\phi \left( g \right)
\quad ,
\label{4.2}
\ee
where $\lambda_h $ are parameters.
On regular lattices,
one usually chooses $\lambda_e = 1 $
and takes $\lambda_h $ for $h \in \NN$ to be negative,
in which case the $\lambda_h $ are called hopping parameters.
The field theory in Eq.\ \bref{4.2} leads to the propagation
of a charged particle through the group lattice.
The hopping parameter $- \lambda_h$ controls
how easily or difficult it is for the charged particle
to propagate over the bond $\left[ hg , g \right]$.

The path integral $Z$ for this system is given by
\be
  Z = \int {{\cal D}\phi {\cal D}\phi^* \exp \left( {-S_0} \right)}
\quad ,
\label{4.3}
\ee
where the measure is
\be
  {\cal D}\phi {\cal D}\phi^* = \prod_{g\in G}
  {{d\phi \left( g \right) d\phi^*\left( g \right)} \over {2\pi }}
\quad .
\label{4.4}
\ee
Free propagation on a group lattice is exactly solvable
if all the irreducible matrix representations of the group $G$
are known
\ct{samuel90a}.
The solution method is based on
the group analog of the Fourier transform.
The result for the action in Eq.\ \bref{4.2} is
\be
  Z = \prod\limits_{ r }
  \left[ {
    \det \left( {
           \sum\limits_{h \in \NN_e}
   \lambda_h D^{ \left( r \right) } \left( h \right)
                } \right)
          } \right]^{ -d_{ r } }
\quad ,
\label{4.5}
\ee
where $D^{\left( r \right)} \left( h \right) $
is the matrix for $h$ in the irreducible representation
$r$ and
$d_{ r }$ is the dimension of the representation.
The product in Eq.\ \bref{4.5} is over
all the irreducible representations $r$ of $G$.
As long as the dimensions of the representations
are not too big,
the determinants which enter in Eq.\ \bref{4.5}
are of moderate size and
can be straightforwardly computed.
For a twisted group lattice,
the matrix in Eq.\ \bref{4.5} is
$$
  \sum\limits_{h\in \NN_e}
   \lambda_h D^{\left( r \right)} \left( h \right)\ =\
   \lambda_e I_{d_{ r} }  +
$$
\be
     \sum\limits_{j=1}^d
  \left( {\lambda_h\Gamma_j^{\left( r \right)}
   \exp \left( {{{2\pi i k_j^{
    \left( r \right)}} \over {L_j}}} \right) +
  \lambda_{h^{-1}}
   \left[ {\Gamma_j^{\left( r \right)}} \right]^{-1}
   \exp \left( {
    {{-2\pi i k_j^{\left( r \right)}} \over {L_j}}
                } \right)
       } \right)
\quad ,
\label{4.6}
\ee
where the $\Gamma_j^{\left( r \right)}$ are the matrices
satisfying Eq.\ \bref{3.3}
in the irreducible representation $r$
and the $k_j^{\left( r \right)}$
are the corresponding spatial momenta.
For the four-dimensional twisted group lattice,
the $\Gamma_j^{\left( r \right)}$ are given
in Sect.\rf{s:cir},
and the product over representations in Eq.\ \bref{4.5} is
\be
    \prod_{r} =
    \prod_{\left\{ {k_{ij}=0} \right\}}^{\left\{ {N_{ij}-1} \right\}}
    \ \prod_{\left\{ {k_i^{\left( r \right)}} \right\} \pmod {E} }
\quad .
\label{4.7}
\ee

Interactions are introduced by considering
products involving more than two fields.
For example,
a ``$\phi$-fourth'' theory is defined
by adding $S_{int}$, given by
\be
  S_{int} =
   \lambda \sum\limits_{g\in G}
   \phi^*\left( g \right) \phi \left( g \right)
   \phi^*\left( g \right) \phi \left( g \right)
\quad ,
\label{4.8}
\ee
to $S_0$,
where $\lambda$ is a coupling constant.
It is straightforward to obtain propagators for this theory
\ct{samuel90a}.
Hence, perturbation theory via Feynman graphs
can be carried out.

Fermionic field theories are obtained
by making $\phi$ anticommuting.
The above equations still hold
with the exception of Eq.\ \bref{4.5} where
the power of the matrix is of the opposite sign, i.\ e.,
one replaces
$ -d_{ r }$ by $ d_{ r } $
in Eq.\ \bref{4.5}.
One could also consider a real scalar field theory.
In this case,
$
  \phi^*\left( {hg} \right) \to \phi \left( {hg} \right)
$
in Eq.\ \bref{4.2},
and the measure in Eq.\ \bref{4.4} becomes
$
  {\cal D}\phi {\cal D} \phi^*
  \to \prod\limits_{g\in G}
     {{d\phi \left( g \right)} \over {\sqrt {2\pi }}}
$.
The solution is Eq.\ \bref{4.5} with
$
  d_{ r } \to d_{ r }/2,
$
and the matrix in Eq.\ \bref{4.6} is the same except
$
  \lambda_h \to
      \left( {\lambda_h + \lambda_{h^{-1}}} \right) / 2
$ and
$
    \lambda_{h^{-1}} \to
      \left( {\lambda_h + \lambda_{h^{-1}}} \right) / 2
$.

\subsection{Lattice Gauge Theories on Group Lattices }
\label{ss:lgtgl}

\hspace{\parindent}
It is straightforward to put gauge theories
on a group lattice,
as soon as the plaquettes in the lattice
are specified.
One selects a gauge group ${\cal G}$.
Frequent choices for ${\cal G}$ are
$
  SU\left( M \right)
$,
$
  U \left( M \right)
$,
and
$
  SO\left( M \right)
$.
One can even use a discrete gauge group,
although it is more difficult or impossible
to take a continuum limit in this case
for certain systems
for which a thermodynamic limit is possible.%
{\footnote{Thermodynamic limits can be considered
for twisted group lattices
by letting $L_i \to \infty$.}}

To put a gauge theory on a lattice,
one uses the standard Wilson method.
For each bond $ \left[ hg , g \right] $ in the lattice,
one assigns a ``link variable''
$U_{hg,g}$ in a matrix representation $R_{\cal G}$
of the gauge group,
i.\ e.,
\be
  U_{hg,g} \in R_{\cal G}
\ , \quad \quad \hbox{ for all } g \in G
                \hbox{ and } h \in \NN
\quad .
\label{4.9}
\ee
Often $R_{\cal G} $ is the fundamental representation.
A link variable, which is oriented in the opposite direction,
is the inverse matrix,
i.\ e.,
\be
  U_{g,hg} = U_{hg,g}^{-1}
\quad .
\label{4.10}
\ee
Usually a unitary representation is used so that
$ U_{hg,g}^{-1} = U_{hg,g}^{\dagger}$.
Equation \bref{4.10} says that a link variable
for a bond using a shift $h^{-1}$
is the inverse of the link variable using a shift $h$
but at the site $h^{-1} g$:
\be
   U_{h^{-1}g, g} =
   U_{hg',g'}^{-1}
\ , \quad \quad
\hbox{ where } g' = h^{-1} g
\quad .
\label{4.11}
\ee
For this reason it is not necessary
to assign link variables
for the bonds associated with inverse elements in $\NN$.
If $h$ is its own inverse, $ h = h^{-1} $,
then the corresponding link variable
must be its own inverse.
Usually there are only discrete choices
for this case and a continuum limit may not be possible.
Such elements $h$ do not occur
for the $d$-dimensional twisted group lattices.

A plaquette $p$ is a series of bonds
that form a closed loop in the lattice.
We denote the set of all plaquettes by ${\cal P}$.
The choice of ${\cal P}$ is at one's disposal.
If $p \in {\cal P}$,
then $p$ is determined
by the starting point $g$
and by a series of elements
$h_1, h_2, \dots , h_m$ in $\NN$,
where $h_m h_{m-1} \cdots h_1 = e$.
The corresponding closed path begins at $g$,
goes to $h_1 g$,
then to $h_2 h_1 g$,
$\dots$,
then to $h_{m-1} h_{m-2} \cdots h_1 g$,
and back to $g$.
Plaquettes which transverse the same path
but start at different points
are considered equivalent.
A plaquette variable $U_p$ is computed
by multiplying the link variables
around the path.
More precisely,
\be
 U_p\ =\
  U_{g, h_{m-1} h_{m-2} \dots h_1 g}
     U_{h_{m-1} h_{m-2} \dots h_1 g ,
                h_{m-2} \dots h_1 g}
 \cdots U_{h_2 h_1 g, h_1 g} U_{h_1 g, g}
\quad .
\label{4.12}
\ee
The gauge theory action $S_{\cal G}$
is then given by
\be
 S_{\cal G} =
   \beta \sum_{p \in {\cal P} }
    \left( {
   Tr \left[ U_p \right] +
   Tr \left[ U_p^{-1} \right]
            } \right)
\quad ,
\label{4.13}
\ee
where $\beta$ is a coupling constant,
and $Tr$ is the matrix trace, i.\ e.,
the character of $U_p$
in the representation $R_{\cal G}$.
The partition function is
\be
 Z_{\cal G} \left( \beta \right) =
   \left( {
      \prod_{b \in {\cal B} } \int dU_b
          } \right)
     \exp ( - S_{\cal G} )
\quad ,
\label{4.14}
\ee
where the measure
involves the product over all bonds in the lattice.
In Eq.\ \bref{4.14},
$dU_b$ is
the left and right group-invariant Haar measure,
normalized so that $\int dU_b = 1$.
The theory governed
by Eq.\ \bref{4.13}
has the gauge invariance
\be
  U_{ hg , g} \to  V_{hg} U_{ hg , g} V_g^{-1}
\quad ,
\label{4.15}
\ee
where $V_g \in R_{\cal G}$.

Matter fields which interact with gauge fields
can be introduced.
Let $\phi$ be a field which transform under the gauge group
via $\phi (g) \to R_{( \phi )} \left[ V_g \right] \phi (g)$,
for some unitary representation $R_{( \phi )}$ of the gauge group.
The action
in Eq.\ \bref{4.2}
is then modified to
\be
  S =  \sum\limits_{g\in G}
  \sum\limits_{h\in \NN_e}
  \lambda_h \phi^*\left( {hg} \right)
    R_{( \phi )} \left[ U_{ hg, g } \right]
             \phi \left( g    \right)
\quad .
\label{4.16}
\ee
It is gauge invariant as long as $U_{ hg , g}$ is transformed
as in Eq.\ \bref{4.15}.
We define $U_{ g, g } = 1$
so that Eq.\ \bref{4.16}
makes sense when $h=e$.

Applying the above construction
to the $d$-dimensional twisted group lattice,
one obtains the following.
There exist link variables
$
  U_{ x_i g,g}
$
for all $i$ and all $g$.
Eq.\ \bref{4.11} for
$  U_{x_i^{-1}g,g} $
gives
$
  U_{x_i^{-1}g,g} = U_{g,x_i^{-1}g}^{-1}
$.
A natural choice
for the plaquette set ${\cal P}$ exists.
On regular hypercubic lattices,
for which all $N_{ij} = 1$,
plaquettes are obtained
by using the elements
$x_i$, $x_j$, $x_i^{-1}$ and $x_j^{-1}$
for any $i<j$.
In other words,
the four corners of a plaquette
in the $i$-$j$ plane are
$ g$, $x_i g$, $x_j x_i g$ and $x_i^{-1} x_j x_i g$.
On a twisted group lattice, however,
such a path is not closed since,
when one moves
from the last site $x_i^{-1} x_j x_i g$
using $x_j^{-1}$,
one does not return to the original site $g$.
Instead, one arrives at
$
  x_j^{-1} x_i^{-1} x_j x_i g = z_{ij} g
$.
On a twisted group lattice,
one must go around the $ij$ square $N_{ij}$
times to return to the original site.
We choose such paths for all $i<j$ to be the plaquettes.
In other words,
the corners of a plaquette are
$$
  p\ \leftrightarrow\
   \left\{ {x_i^{-1}x_jx_iz_{ij}^{N_{ij}-1}g,
    x_jx_iz_{ij}^{N_{ij}-1}g,
    x_iz_{ij}^{N_{ij}-1}g,z_{ij}^{N_{ij}-1}g, \ldots ,
    } \right.
$$
\be
   \left. {
   x_iz_{ij}g ,z_{ij}g, x_i^{-1}x_jx_ig, x_jx_ig, x_ig, g} \right\}
\quad .
\label{4.17}
\ee
The set ${\cal P}$ for the $d$-dimensional twisted group lattice
is obtained from Eq.\ \bref{4.17}
by using all $g$ and all $i<j$.
The action and partition function are then
given as in Eqs.\ \bref{4.13} and \bref{4.14}.

Non-trivial twisted group lattices,
for which some $N_{ij} >1$,
possess
link gauge invariances.
The situation is similar to the adjoint action
$
 S_{\cal G} =
   \beta_A \sum_{p \in {\cal P} }
    \left( {
   Tr\left[ U_p \right] Tr \left[ U_p^{\dagger} \right]
            } \right)
$
of an $SU ( M )$ lattice gauge theory
on a regular lattice.
If $U_p \to V U_p$ for any $V$ in the center $Z_M$
of $SU ( M )$ then $S_{\cal G}$ is left unchanged.
This is a local symmetry
since the transformation can be performed on any link.
Hence, the effective gauge group is $SU ( M ) / Z_M$.
Using the adjoint action
in lieu of the Wilson action,
which is based on the fundamental representation,
has a dramatic effect on the physics.
A non-overlapping Wilson loop
in the fundamental representation vanishes.
This result follows from
the above local link symmetry.

To determine the additional local invariances
of the gauge theory on a $d$-dimensional twisted group lattice,
it is useful to define some concepts.
The center ${\cal Z}_G$ of the group $G$
governing the group lattice
in Eq.\ \bref{1.1}
is generated by the $z_{ij}$ elements, i.\ e.,
\be
  {\cal Z}_G\ =\
     \left\{ {
     \prod_{i<j}
     z_{ij}^{n_{ij}} \ , \hbox{ such that }
    n_{ij}=0,1,\ldots ,N_{ij}-1
              } \right\}
\quad .
\label{4.18}
\ee
Let us say that $g$ is {\it above or below} $g^{\prime}$
if $g = z g^{\prime}$ for some $z \in {\cal Z}_G$.
A bond $\left[ {x_ig,g} \right]$
is {\it above or below}
the bond $\left[ {x_i g^{\prime}, g^{\prime}} \right]$
if $g$ is above or below $g^{\prime}$.
Let $g$ be a fixed point on the group lattice
and fix $i$.
Consider the transformation
\be
  U_{x_i zg, zg } \to V_z U_{x_i zg, zg}
\quad ,
\label{4.19}
\ee
for $V_z$ in the center
${\cal Z}_{{\cal G}}$
of the gauge group representation $R_{\cal G}$.
Equation \bref{4.19}
is a local transformation
for all link variables on the bonds above or below
the bond $\left[ x_i g, g \right]$.
Let
$ {\cal I} $
denote the identity matrix
in the representation $R_{\cal G}$.
If for every $z \in {\cal Z}_G$,
one has
$$
  \prod_{n=0}^{N_{ij}-1}
  V_{z_{ij}^n z} = {\cal I}
\ , \quad  \hbox{ for every } j > i
\quad  (i \hbox{ fixed})
\quad ,
$$
\be
  \prod_{n=0}^{N_{ji}-1}
  V_{z_{ji}^n z} = {\cal I}
\ ,  \quad \hbox{ for every } j < i
\quad  (i \hbox{ fixed})
\quad ,
\label{4.20}
\ee
then the plaquette variables $Tr U_p$
are left unchanged,
so that the transformation
in Eq.\ \bref{4.19}
is a local symmetry.
For $d=2$,
the condition
in Eq.\ \bref{4.20}
simplifies to
\be
  \prod_{n=0}^{N_{12}-1} V_{z_{12}^n} = {\cal I}
\quad .
\label{4.21}
\ee
It is easy to find solutions to
Eq.\ \bref{4.21}.
For example,
$
  U_{x_1, e} \to VU_{x_1, e}
$,
$
  U_{z_{12} x_1, z_{12}} \to V^{-1} U_{ z_{12} x_1, z_{12}}
$,
where $V \in {\cal Z}_{{\cal G}}$,
produces a local link symmetry.

The behavior of gauge theories
on four-dimensional twisted group lattices
is an interesting topic which goes
beyond the current work
and may require the use of computer simulations.
Important issues are the question of confinement,
asymptotic freedom,
and the continuum limit.
Due to the local link invariance,
the nature of confinement,
if it exists,
might be different
in the sense that certain Wilson loops
may have zero expectation value.
The link gauge symmetry may be broken, if so desired,
by introducing additional or different plaquettes
in ${\cal P}$.
It can also be broken by introducing matter fields
which couple to the gauge fields as
in Eq.\ \bref{4.16}.

Two-dimensional gauge theories
on {\it regular\/} lattices are exactly solvable.
The strong coupling expansion can be computed
and gives results valid for all values of the coupling.
On each plaquette,
one performs the following character expansion
\be
  \exp \left[ {\beta Tr\left( {U_p} \right) +
               \beta Tr\left( {U_p^{\dagger} } \right)} \right] =
    {\cal Z}_0\left( {\beta } \right)
      \sum\limits_{\left( \rho \right)}
         d_{\rho}
         z_{\rho}\left( {\beta } \right)
     \chi_{\rho}\left( {U_p} \right)
\quad ,
\label{4.22}
\ee
where the sum is over all irreducible representations
$\left( \rho \right)$ of the gauge group ${\cal G}$.
In Eq.\ \bref{4.22},
$\chi_{\rho}$ and
$d_{\rho}$ denote
the character and dimension of $\rho$.
The single plaquette partition function
${\cal Z}_0\left( {\beta } \right)$
is given by
\be
 {\cal Z}_0\left( {\beta } \right) =
 \int dU
  \exp \left[ {\beta Tr\left( {U} \right) +
               \beta Tr\left( {U^{\dagger} } \right)} \right]
\quad .
\label{4.23}
\ee
Explicit formulas for ${\cal Z}_0\left( {\beta } \right)$ and
the character coefficients
$ z_{\rho}\left( {\beta } \right)$
can be found in many places.
See, for example,
Ref.\ct{characters}.
After the expansion
in Eq.\ \bref{4.22}
is performed,
integration over link variables can be explicitly done using
the integral formulas
$$
 \int {dU} \chi_{\rho} \left( {AU} \right)
           \chi_{\sigma} \left( {U^{\dagger} B} \right) =
    {{\delta_{\rho\sigma}}
       \over {d_{\rho}}}
   \chi_{\rho}\left( {AB} \right)
\quad ,
$$
\be
  \int {dU} \chi_{\rho}\left( {AUBU^{\dagger} } \right) =
     {{1}
     \over {d_{\rho}}}
   \chi_{\rho}\left( A \right)
   \chi_{\rho}\left( B \right)
\quad ,
\label{4.24}
\ee
which follow from the orthogonality relations
for integrals over irreducible representations.
Only two characters appear
in performing the strong coupling integrations
because each bond is shared by only two plaquettes
for a two-dimensional surface.
The final result is
\be
  Z_{{\cal G}} \left( {\beta } \right) =
   \left[ {{\cal Z}_0\left( {\beta } \right)} \right]^V
   \sum\limits_{\left( \rho \right)}
           d_{\rho}^{2-2g}
   \left[ {z_{\rho} \left( {\beta } \right)} \right]^V
\quad ,
\label{4.25}
\ee
where $V = L_1 L_2$ is the volume of the system,
i.\ e., the number of plaquettes,
and $g$ is the genus of the surface.

Gauge theories
on {\it twisted\/} two-dimensional group lattices
are also exactly solvable.
With the above definition of ${\cal P}$,
a two-dimensional twisted group lattice
is a closed two-dimensional surface since each bond
is shared by exactly two plaquettes.
Gluing plaquettes together
at such bonds produces the two-dimensional surface.
Since Eq.\ \bref{4.25}
holds for an arbitrary two-dimensional lattice,
one can compute the genus of the periodic twisted group lattice
using lattice gauge theory.
Orient the lattice so that the $x$ and $y$ axes
point respectively in the horizontal and vertical directions.
Consider the plaquettes along a horizontal strip
of length $N_{12}$.
Perform the integrations of the link variables
on all interior vertical bonds.
There are $N_{12} \left( N_{12}-1 \right)$
such bonds.
These integrations yield a factor of
$d_{\rho}^{ - N_{12} \left( N_{12}-1 \right) }$
via Eq.\ \bref{4.24}
and produce $N_{12}$ separate Wilson-type loops.
One can do this for all strips of size $1 \times N_{12}$.
The result is $N_{12}$ separate gauge theories
on regular toroidal lattices
with spatial volumes of $V/N_{12}$.
The remaining integrations then proceed
as for the regular lattice case.
Note that there are also factors of
$
   \left[ {
     z_{\rho} \left( {\beta } \right)
          } \right]^{V}
$
and
$\left[ {{\cal Z}_0\left( {\beta } \right)} \right]^{V}$
from Eq.\ \bref{4.22},
where ${V}$ is the space-time volume:
${V} = L_1 L_2$.
Hence, one finds
\be
  Z_{N_{12}}\left( {\beta } \right) =
  \left[ {{\cal Z}_0\left( {\beta } \right)} \right]^{V}
  \sum\limits_{\left( \rho \right)}
   d_{\rho}^{\left( {1-N_{12}} \right)V}
   \left[ {z_{\rho}\left( {\beta } \right)} \right]^{V}
\quad .
\label{4.26}
\ee
One concludes that the genus $g$
is
\be
  g = {{\left( {N_{12}{-}1} \right) V } \over 2} + 1
\quad .
\label{4.27}
\ee
The genus is generally quite large
and grows with the space-time volume.
A two-dimensional twisted group lattice is indeed quite twisted.
When $N_{12} = 1$,
one recovers the ordinary lattice-gauge-theory result on the torus.

Equation \bref{4.27}
can be computed using Euler's theorem
which says that
$2 (1-g) = v - e + f$,
where $v$, $e$ and $f$ are
respectively the number of vertices, edges and faces
of the lattice.
For the twisted group lattice,
$v= N_{12} V$, $e = 2 N_{12} V$
and $f = V$.

``Planar'' Wilson loops can be evaluated.
They either vanish due to the local link symmetry
or have area law.
Confinement is a property of gauge theories
on two-dimensional twisted group lattices.

\subsection{Automatic Dirac-Gamma-Matrix Structure}
\label{ss:adgms}

\hspace{\parindent}
It was noted
in Ref.\ct{samuel92a}
that the $\{ k_{ij} = 1 \}$ sector of the
$\{ N_{ij} = 2 \}$ case
leads to Dirac-gamma-matrix structure.
If anti-commuting fields are permitted to propagate
on this twisted group lattice,
then Dirac fermions emerge.
It is interesting to determine whether these fermions
are naive, staggered, or of some other type.

To analyze the situation,
it is convenient to represent
the $d$-dimensional twisted group lattice
as a $d ( d + 1 ) / 2$-dimensional lattice
in which there are screw dislocation defects
at the centers of all plaquettes
\ct{samuel91b,samuel92a}.
One starts with
the $d ( d + 1 ) / 2$-dimensional periodic lattice
\be
 L_1 \times L_2 \times \ldots \times L_d \times
 N_{12} \times N_{13} \times \ldots \times N_{d-1d}
\quad .
\label{4.28}
\ee
Let
\be
  {\bf e}_1,    {\bf e}_2,    \ldots , {\bf e}_d,
  {\bf e}_{12}, {\bf e}_{13}, \ldots , {\bf e}_{d-1d}
\quad ,
\label{4.29}
\ee
be the standard orthonormal basis.
The first $d$ basis vectors
${\bf e}_1,    {\bf e}_2,    \ldots , {\bf e}_d$
correspond to the usual $d$ directions.
There are plaquettes in each of the $d ( d - 1 ) / 2$ planes.
One arbitrarily appends $d ( d - 1 ) / 2$ directions
associated with these planes.
The remaining $d ( d - 1 ) / 2$ vectors
$  {\bf e}_{12}, {\bf e}_{13}, \ldots , {\bf e}_{d-1d} $
are a basis for this auxiliary space.
A point ${\bf y}$ in ordinary space corresponds to
\be
  {\bf y} = \sum\limits_{i=1}^d {y_i {\bf e}_i}
\quad .
\label{4.30}
\ee
A point in the full space is denoted by $( {\bf y}, {\bf w})$ where
\be
  {\bf w} = \sum\limits_{1 \le i<j\le d}^{} {w_{ij} {\bf e}_{ij}}
\quad ,
\label{4.31}
\ee
that is,
\be
  \left( {{\bf y},{\bf w}} \right) =
  \sum\limits_{i=1}^d {y_i{\bf e}_i} +
  \sum\limits_{1\le i<j\le d}^{} {w_{ij}{\bf e}_{ij}}
\quad .
\label{4.32}
\ee

Screw dislocations are introduced by modifying
the direction of bonds in the $i$-direction,
for $i=1, \dots , d-1$.
Bonds exist between
$\left( {{\bf y},{\bf w}} \right)$
and
$  \Bigl( {{\bf y}{+} {\bf e}_i,
  {\bf w} {-} \sum\limits_{j=i+1}^d {{\bf e}_{ij}y_j}} \Bigr)
$.
A movement along this bond is associated with $x_i$.
A movement between
$\left( {{\bf y},{\bf w}} \right)$
and
$
  \left( {{\bf y},{\bf w} {+} {\bf e}_{ij}} \right)
$
is associated with $z_{ij}$.
It is easy to check that, for $j>i$,
an $x_i$ movement,
followed by an $x_j$ movement is
an $x_j$ movement,
followed by an $x_i$ movement,
followed by a $z_{ij}$ movement.
Noting that we order ``movements'' from right to left,
the last statement corresponds to a physical realization
of the equation $x_j x_i = z_{ij}  x_i x_j $ for $j>i$.

Let $\psi$ be an anticommuting field and
let $\bar \psi$ be its conjugate.
Consider the following action
$$
  A \ =\ \sum\limits_{{\bf y},{\bf w}} {}\sum\limits_{i=1}^d {}
   \left( {\lambda_{{\bf e}_i}
        \bar \psi \Bigl( {{\bf y}{+}{\bf e}_i, {\bf w} {-}
     \sum\limits_{j=i+1}^d {{\bf e}_{ij}y_j}} \Bigr) }
     \right. \psi \left( {{\bf y}, {\bf w}} \right)\ +
$$
\be
  \left. { \lambda_{-{\bf e}_i}
   \bar \psi \Bigl( {{\bf y}{-}{\bf e}_i,
  {\bf w}{+}\sum\limits_{j=i+1}^d {{\bf e}_{ij}y_j}} \Bigr)
   \psi \left( {{\bf y},{\bf w}} \right) } \right)\ -\
    \sum\limits_{{\bf y},{\bf w}} \lambda_0
   \bar \psi \left( {{\bf y},{\bf w}} \right)
        \psi \left( {{\bf y},{\bf w}} \right)
\quad ,
\label{4.33}
\ee
where $\lambda_{{\bf e}_i}$ and $\lambda_{-{\bf e}_i}$
are hopping parameters for the shifts
associated with $x_i$ and $x_i^{-1}$.
The parameter $\lambda_0$,
which is $\lambda_h$ for $h=e$,
is a chemical potential or a mass.
Note that the arguments in $\bar \psi$
correspond to the shifts created by $x_i$, $x_i^{-1}$ or $e$.

Perform a Fourier transformation
with respect to the $d (d-1) /2 $ variables $ \{ w_{ij} \}$ via
$$
  \psi \left( {{\bf y},{\bf w}} \right)\ =\
  \left( {\prod_{1\le i<j\le d}^{} {}
  {1 \over {\sqrt {N_{ij}}}}
    \sum\limits_{k_{ij}=0}^{N_{ij}-1} {}} \right)
   \exp \left[ {2\pi i\sum\limits_{1\le i<j\le d} {}
    {{k_{ij}w_{ij}} \over {N_{ij}}}} \right]
    \tilde \psi \left( {{\bf y},
     \left\{ {k_{ij}} \right\}} \right)\quad,
  \qquad\qquad\phantom{.}
$$
\be
  \bar \psi \left( {{\bf y},{\bf w}} \right)\ =\
  \left( {\prod_{1\le i<j\le d}^{} {}{1 \over {\sqrt {N_{ij}}}}
   \sum\limits_{k_{ij}=0}^{N_{ij}-1} {}} \right)
   \exp \left[ {-2\pi i\sum\limits_{1\le i<j\le d} {}
    {{k_{ij}w_{ij}} \over {N_{ij}}}} \right] \tilde {\bar \psi}
   \left( {{\bf y},\left\{ {k_{ij}} \right\}} \right)
\quad ,
\label{4.34}
\ee
where the transformed fields are denoted by a tilde.
The action becomes
$$
  A \ =\
   \sum\limits_{ \left\{ {k_{ij}=0}
                \right\} }^{ \left\{ N_{ij}-1 \right\} }
   \sum\limits_{{\bf y}}
   \sum\limits_{i=1}^d {}\left( {\lambda_{{\bf e}_i}
    \alpha_i
  \tilde {\bar \psi} \left( {{\bf y}{+}{\bf e}_i,
         \left\{ {k_{ij}} \right\}} \right)} \right.
    \tilde \psi \left( {{\bf y},\left\{ {k_{ij}} \right\}} \right) +
$$
\be
  \left. {\lambda_{-{\bf e}_i}
   \alpha_i^{-1}
 \tilde {\bar \psi} \left( {{\bf y}{-}{\bf e}_i,
         \left\{ {k_{ij}} \right\}} \right)
       \tilde \psi \left( {{\bf y},
         \left\{ {k_{ij}} \right\}} \right)} \right) -
    \sum\limits_{{\bf y},\left\{ {k_{ij}} \right\}}
   \lambda_0
 \tilde {\bar \psi}
    \left( {{\bf y},\left\{ {k_{ij}} \right\}} \right)
        \tilde \psi
    \left( {{\bf y},\left\{ {k_{ij}} \right\}} \right)
\quad ,
\label{4.35}
\ee
where
\be
 \alpha_i \equiv e^{2\pi i\sum\limits_{j=i+1}^d
  {{k_{ij}y_j} \over {N_{ij}}}}
\quad .
\label{4.36}
\ee
The theory has factorized into sectors
corresponding to the $\{ k_{ij} \}$.

Consider the case $N_{ij}=2$ and the subsector $ k_{ij} = 1 $
for all $1 \le i < j \le d$
and let
$
   \lambda_{{\bf e}_i} = - \lambda_{-{\bf e}_i} = {i \over 2}
$
for $i=1, 2, \ldots , d $.
The phase factors $\alpha_i$
in Eq.\ \bref{4.26} become
\be
     \alpha_i  \to  \left( {-1} \right)^{\sum\limits_{j=i+1}^d {y_j}}
\ , \quad \quad
 \alpha_i^{-1}  \to  \left( {-1} \right)^{\sum\limits_{j=i+1}^d {y_j}}
\quad .
\label{4.37}
\ee
Let $\chi$ be
the ${\left\{ {k_{ij}=1} \right\}}$ Fourier component
of $\psi$:
\be
  \chi \left( {\bf y} \right) =
   \tilde \psi \left( {{\bf y},
        \left\{ {k_{ij}=1} \right\}} \right)
\ , \quad \quad
      \bar \chi \left( {\bf y} \right) =
   \tilde {\bar \psi} \left( {{\bf y},
        \left\{ {k_{ij}=1} \right\}} \right)
\quad .
\label{4.38}
\ee
The action $A_{\left\{ {k_{ij}=1} \right\}}$ in this sector is
\be
  A_{\left\{ {k_{ij}=1} \right\}} =
   \sum\limits_{\bf y} {}\sum\limits_{i=1}^d {}
   {i \over 2} \alpha_i
   \left( {\bar \chi \left( {{\bf y}{+}{\bf e}_i} \right)
   \chi \left( {\bf y} \right)-\bar \chi
   \left( {{\bf y}{-}{\bf e}_i} \right)
   \chi \left( {\bf y} \right)} \right) -
   \lambda_0 \sum\limits_{\bf y} {}\bar \chi
   \left( {\bf y} \right)\chi \left( {\bf y} \right)
\ ,
\label{4.39}
\ee
where
$\alpha_i = \left( {-1} \right)^{\sum\limits_{j=i+1}^d {y_j}} $
is a sign factor appropriate
for Kogut-Susskind staggered fermions
\ct{ks}.
Hence,
the $\{ k_{ij} = 1 \}$ sector of the $\{ N_{ij} = 2 \}$
twisted group lattice contains staggered fermions.

\subsection{Space-Time Discrete Symmetries}
\label{ss:stds}

\hspace{\parindent}
Regular hypercubic lattices
have discrete space-time symmetries,
such as translations by the lattice spacing
and $90^\circ$ rotations in a plane.
The same is true
for the twisted $d$-dimensional group lattices.
Let $g'$ be a fixed element.
Field theories on group lattices
possess the symmetry
\ct{samuel90a}
\be
  g \to gg' \quad\hbox{ for all }g
\quad .
\label{4.40}
\ee
For example, in the case of a scalar field, $\phi$,
the transformation $\phi (g) \to \phi (gg')$ leaves
the action invariant
because the transformation
in Eq.\ \bref{4.40}
maintains the bond structure:
if $g_1$ and $g_2$ are
nearest neighbor sites,
then $g_1 g'$ and $g_2 g'$ are also
nearest neighbor sites.
Note that one must multiply $g'$ from the right
since nearest neighbor sites are determined by
multiplication from the left.
A translation in the $i$th direction by $m_i$
lattice spacings corresponds to
Eq.\ \bref{4.40} for $g' = x_i^{m_i}$,
i.\ e.,
$
  g \to g x_i^{m_i}
$.
When an element is represented as
in Eq.\ \bref{1.1},
$m_i$ is added to the exponent $n_i$ of $x_i$
while leaving the exponents of the other $x_j$ unchanged.
Hence
$
  g \to g x_i^{m_i}
$
corresponds to a shift $n_j \to n_j + \delta^i_j m_i$.
Of course, $z_{ij}$ exponents may change,
but this is a necessary consequence of twisting.
One also has translations in the ``extra dimensions''
via
$
  g \to g z_{ij}^{m_{ij}}
$,
but these can be generated by
translations in $i$ and $j$ directions
using
$
  x_j^{-1} x_i^{-1} x_j x_i
$
repeatedly $m_{ij}$ times.
The translation group of the twisted lattice
is different from the one of the regular hypercubic lattice
because of twisting.
It is precisely this difference
which makes the twisted group lattice interesting.

Regular hypercubic lattices possess
$90^\circ$ rotational symmetries.
Let $R_{ij}$ denote a rotation of the $ij$-plane by $90^\circ$.
Representing the regular lattice as an abelian group lattice,
one defines $R_{ij}$ by
$$
  R_{ij}\left[ e \right] = e
\ , \quad \quad
  R_{ij}\left[ {x_i} \right] = x_j
\quad ,
$$
\be
  R_{ij}\left[ {x_j} \right] = x_i^{-1}
\ , \quad \quad
  R_{ij}\left[ {x_k} \right] = x_k
\quad ,
\label{4.41}
\ee
where $i$, $j$, and $k$ are distinct,
and
where we denote the action of $R_{ij}$
on a point $g$ by $R_{ij} \left[ g \right]$.
On a general element,
$R_{ij}$ is determined
from Eq.\ \bref{4.41}
and the requirement that it be a homomorphism
\be
 R_{ij} \left[ g_1 g_2 \right] =
 R_{ij} \left[ g_1 \right]  R_{ij} \left[ g_2 \right]
\ , \quad \quad
 R_{ij} \left[ g^{-1} \right] =
 \left(  R_{ij} \left[ g \right] \right)^{-1}
\quad .
\label{4.42}
\ee

Discrete rotations satisfy the group relations
$$
  R_{ij} R_{ji} = E
\ , \quad \quad
  \left( {R_{ij}} \right)^4 = E
\quad ,
$$
\be
  R_{jk} R_{ki} R_{kj} = R_{ij}
\ , \quad \quad
  R_{ki} R_{kj} R_{ik} = R_{ij}
\ , \quad \quad
  R_{ij} R_{kl} = R_{kl}R_{ij}
\quad ,
\label{4.43}
\ee
where $i$, $j$, $k$ and $l$ are distinct,
and
$E$ is the identity transformation.
Equation \bref{4.43} is the lattice analog
of the continuum $O ( d ) $ Lie group relations.

The twisted $d$-dimensional group lattices also possess
discrete rotational symmetries
if the lengths $L_i$ in all directions are equal: $L_i =L$,
and the twistings are all equal: $N_{ij} = N$
for $i<j$.
We define the action $ R_{ij} $ on group elements
via Eqs.\ \bref{4.41} and \bref{4.42}%
{\footnote {It is important
to maintain the order
in Eq.\ \bref{4.42}
since now the group
is non-abelian.}}
and by requiring $ R_{ij} $ to respect
the group relations.
This implies the additional actions
$$
  R_{ij}\left[ {z_{ij}} \right] = z_{ij}
\ , \quad \quad
  R_{ij}\left[ {z_{kl}} \right] = z_{kl}
\quad ,
$$
$$
  R_{ij}\left[ {z_{ik}} \right] = z_{jk}
\ , \quad \quad
  R_{ij}\left[ {z_{ki}} \right] = z_{kj}
\quad ,
$$
\be
  R_{ij}\left[ {z_{jk}} \right] = z_{ik}^{-1}
\ , \quad \quad
  R_{ij}\left[ {z_{kj}} \right] = z_{ki}^{-1}
\quad ,
\label{4.44}
\ee
where $i$, $j$, $k$ and $l$ are distinct, and
where, for notational convenience, we have defined
$
  z_{ji} \equiv z_{ij}^{-1} \hbox{ for } j>i
$.
The group relations are all satisfied:
Since
$
  e = x_l^L
$,
one needs
$
  e = R_{ij}\left[ e \right] \mathop=\limits^?
   R_{ij}\left[ {x_i^L} \right] = x_j^L
    \mathop=\limits^\surd e
$,
$
   e \mathop=\limits^?
    R_{ij}\left[ {x_j^L} \right] = x_i^{-L}
     \mathop=\limits^\surd e
$,
and
$
   e \mathop=\limits^?
    R_{ij}\left[ {x_k^L} \right] = x_k^L
     \mathop=\limits^\surd e
$.
These equations show why all $L_i$ must be equal.
Likewise, the relation
$
  x_j^{-1} x_i^{-1} x_j x_i = z_{ij}
$
is respected by rotations
if Eq.\ \bref{4.44} holds.
As long as
$
  N_{ij} = N
$ for all $i<j$,
$
  z_{ij}^{N_{ij}} = e
$
is preserved by rotations.

Reflections can be defined
in a manner analogous to rotations.
The reflection of the $i$th coordinate corresponds to
an operator $ R_i$ satisfying
$$
  R_i \left[ x_i \right] = x_i^{-1}
\ , \quad \quad
   R_i \left[ x_j \right] = x_j
\quad ,
$$
\be
  R_i \left[ z_{ij} \right] = z_{ij}^{-1}
\ , \quad \quad
  R_i \left[ z_{ji} \right] = z_{ji}^{-1}
\ , \quad \quad
  R_i \left[ z_{jk} \right] = z_{jk}
\quad ,
\label{4.45}
\ee
where $i$, $j$ and $k$ are distinct.

The discrete rotations and reflections are
symmetries of scalar field theories on twisted group lattices
because the bond structure is maintained.
Of course,
the couplings $\lambda_h$
for $h \ne e$
in Eq.\ \bref{4.2}
must all be
equal:
$
  \lambda _{x_i} = \lambda _{x_i^{-1}} = \lambda
$.
Likewise,
the gauge theory defined
in Eq.\ \bref{4.13}
is invariant under discrete rotations and reflections
when
\be
 U_{hg,g} \to U_{R_{ij}[hg],R_{ij}[g]}
\ , \quad \quad \hbox{ or } \quad
 U_{hg,g} \to U_{R_{i}[hg],R_{i}[g]}
\quad .
\label{4.46}
\ee
In Eq.\ \bref{4.46},
the direction of the link variable
is rotated or reflected.

The group of rotations (and reflections)
on twisted $d$-dimensional group lattices
is isomorphic to the full point group of
the corresponding hypercubic lattices.
In other words,
the relations in Eq.\ \bref{4.43}
hold for the twisted-lattice case as well.

\section{Conclusions}
\label{s:c}

\hspace{\parindent}
In this work,
we have solved the $4$-dimensional twisted group lattice
by finding all the irreducible matrix representations.
A key step was prime factorization.
Given the irreducible representations,
propagators for free field theories can be computed
and a perturbation series for an interacting theory
can be developed.

In addition,
we have shown how to define field theories
on group lattices.
Scalar interactions are introduced by including
polynomials in fields
of order three or higher in the action.
Gauge theories are obtained \`a la Wilson
by using link variables.
Likewise,
fermions can be introduced.
We have shown that,
for the $\{ N_{ij}=2 \}$ system,
staggered-gamma-matrix structure arises naturally
in one sector.
In summary,
the situation for
twisted group lattices is as good as for regular lattices.
The difference between the two lattices
is that the former involves
a non-trivial space-time structure.

As shown in Sect.\rf{ss:adgms},
the $4$-dimensional twisted group lattice
can be viewed as a non-trivial compactification
of a ten-dimensional lattice.
The lattice is not even locally a product
of a $4$-dimensional lattice
with a $6$-dimensional lattice.
It is perhaps curious that
superstring theory is naturally formulated in ten dimensions.
It would be interesting to develop a compactification
of a ten-dimensional continuum theory
with non-trivial-holonomy aspects
similar to the $4$-dimensional twisted group lattices.
One can adopt the opposite viewpoint:
The twisted lattice can be used as a model
for discrete ten-dimensional compactification.
A continuum limit can be obtained
by taking the lengths $L_i$ to infinity
and approaching a critical point,
as one does for a regular lattice.%
{\footnote{One
can also consider taking the $N_{ij}$ to infinity
to obtain a continuum limit in the internal space.}}

One can exploit the Kaluza-Klein analogy
as a means of establishing contact
with low-energy phenomenology.
Indeed,
continuum compactification models contain higher-mass
Kaluza-Klein modes.
These modes are usually heavy
and do not effect low-energy physics.
Likewise,
the $4$-dimensional twisted group lattice
has sectors with $\{ k_{ij} \ne 0 \}$,
which are the lattice analogs of Kaluza-Klein modes.
If these modes are heavy,
then the low-energy theory
is governed by the $\{ k_{ij} = 0 \}$ sector.
This sector is equivalent to an $\{ N_{ij} = 0 \}$ model,
which is an ordinary $4$-dimensional hypercubic lattice.
Hence,
if the Kaluza-Klein modes are heavy
for a field theory on a twisted group lattice,
then the low-energy physics should be similar
to the corresponding field theory
on a regular lattice.
We have checked this idea
for a free scalar field on the $\{ N_{ij} = 2 \}$ lattice.
We find that the $\{ k_{ij} = 0 \}$ sector has the lowest mass.
In a continuum limit,
the $\{ k_{ij} \ne 0 \}$ modes become heavy,
as expected.

We have also demonstrated that the twisted group lattice
possesses discrete translational and rotational invariances.
For regular lattices,
this is usually sufficient to ensure full Euclidean symmetry
in a continuum limit.
The same is likely to be true for twisted group lattices,
although future work is needed to verify this.
In contrast,
for a non-trivial quantum plane,
defined by the relations
$x_j x_i = q_{ij} x_i x_j$,
for $i<j$,
there are no discrete rotational symmetries
analogous to Eq.\ \bref{4.41},
unless the space-time dimension is two
or unless $q_{ij} = -1$ for all $i<j$.
This is easily checked by explicit computation.
It also can be seen by exploiting the magnetic-field analogy
\ct{samuel91b,samuel92a}.
Propagation on a quantum plane is equivalent
to a particle of charge $e$ moving on a regular lattice
in the presence of field strengths $F_{ij}$ given by
$F_{ij} = -i \ln \left( { q_{ij} } \right) / (e a^2)$,
where $a$ is the lattice spacing.%
{\footnote{The field strengths
are real only when the $q_{ij}$ are phases,
so that the discussion should be restricted to this case.}}
When such a charged particle moves around a plaquette
in the $ij$ plane,
its wave function is multiplied by
$ \exp \left( {i e a^2 F_{ij} } \right) = q_{ij}$.
Since the total magnetic field points in some arbitrary direction,
discrete Euclidean rotation invariance is broken.
If discrete Euclidean rotation is to be achieved,
a more general approach is needed.

{}From the twisted-group-lattice viewpoint,
one can understand the difficulties of defining
interacting field theories for quantum hyperplanes.
A quantum plane system in the $ij$ direction, with
$q_{ij}$ being a rational phase,
corresponds to a sector
of the free twisted group lattice
with $ k_{ij}$ determined by
$q_{ij} = \exp \left( {2 \pi i k_{ij}/N_{ij} } \right)$,
i.\ e.,
Kaluza-Klein modes
with internal lattice momenta of
$ 2 \pi k_{ij}/ N_{ij} $.
In a free theory,
such modes are decoupled.
However,
when interactions,
which involve the product of three or more fields,
are present,
those modes couple to each other.
One needs to include the other $ k_{ij}$ momenta
and, hence, other $q_{ij}$ values.
For this reason
it is probably difficult to define
interacting systems for quantum planes.
Twisted group lattices are,
in some sense,
the natural generalization of quantum planes.
For twisted group lattices,
discrete Euclidean invariances are achieved
and interacting field theories are not problematic.

An interesting question is the behavior
of non-abelian gauge theories on twisted group lattices.
Since non-perturbative effects are expected to be important,
one must resort to numerical methods
such as Monte Carlo simulations.
By viewing the $d$-dimensional twisted group lattice
as a particular $d + d(d-1)/2$ lattice,
computer simulations can be performed.

In short,
there is much to understand about twisted group lattices,
both as a model of discrete compactification
and as a possible deformation of space-time structure.

\medskip
\medskip

\noindent
{\large \bf Acknowledgements}

We thank
J.\ Wess
for brief discussions.
This work is supported in part
by a NATO Collaborative Research Grant
(grant number CRG 930761),
by the United States Department of Energy
(grant number DE-FG02-92ER40698),
by the Alexander von Humboldt Foundation,
and by the PSC Board of Higher Education at CUNY.

\vfill\eject

\appendix
\noindent
{\Large \bf Appendix:
\large \bf Proof of Irreducibility and Completeness}
\label{s:A}
\vskip .5cm
\endsecteqno
\secteqnoA

To prove irreducibility and completeness of the representations,
it suffices
\ct{hamermesh}
to show that
\be
  \sum\limits_{\left( r \right)} {d_r^2} =
    o \left( G \right)
\quad ,
\label{A.1}
\ee
and that
\be
 \sum\limits_{g\in G}
 {\chi^{\left( r \right)} \left( g \right)
  \chi^{\left( s \right)}\left( {g^{-1}} \right)} =
  \delta^{\left( r \right) \left( s \right)}
  o\left( G \right)
\quad ,
\label{A.2}
\ee
which expresses the orthogonality of characters.
Here, $\chi^{\left( r \right)}$ denotes
the character of the representation $r$.
The character is obtained by taking the trace
of the matrix.
Equation \bref{A.1}
was already obtained
in Eq.\ \bref{3.38},
so that it is only necessary to prove
Eq.\ \bref{A.2}.

Express the element $g$
in Eq.\ \bref{A.2} as
\be
  g = x_1^{n_1} x_2^{n_2} x_3^{n_3} x_4^{n_4}
      z_{12}^{n_{12}} z_{13}^{n_{13}} z_{14}^{n_{14}}
      z_{23}^{n_{23}} z_{24}^{n_{24}} z_{34}^{n_{34}}
\quad .
\label{A.3}
\ee
Recall that a representation is determined by the integers
$$
  \left\{ {k_{12}, k_{13}, k_{14},
           k_{23}, k_{24}, k_{34},
     k_1, k_2, k_3, k_4} \right\}
\quad .
$$
Since two representations $r$ and $s$ appear
in Eq.\ \bref{A.2},
we use a superscript to distinguish
quantities associated with $r$ and $s$.
For example,
the representation $r$ corresponds to
\be
     r \ \leftrightarrow \
  \left\{ { k_{12}^{\left( r \right)}, k_{13}^{\left( r \right)},
            k_{14}^{\left( r \right)}, k_{23}^{\left( r \right)},
            k_{24}^{\left( r \right)}, k_{34}^{\left( r \right)},
            k_1^{\left( r \right)}, k_2^{\left( r \right)},
            k_3^{\left( r \right)}, k_4^{\left( r \right)}
          } \right\}
\quad .
\label{A.4}
\ee

Using Eqs.\ \bref{3.1} and \bref{A.3},
it follows that
\be
  \chi^{\left( r \right)}\left( g \right)
  \chi^{\left( s \right)}\left( {g^{-1}} \right)
  \  \propto\
     \exp \left[ {
        2\pi i\sum\limits_{i<j}
          {{{\Delta k_{ij}n_{ij}} \over {N_{ij}}}}
                  } \right]
\quad ,
\label{A.5}
\ee
where
$
  \Delta k_{ij} \equiv
     k_{ij}^{\left( r \right)} - k_{ij}^{\left( s \right)}
$.
Hence, the sum over the $n_{ij}$
leads to the conclusion that
\be
  \sum\limits_{n_{ij}}
   {\chi^{\left( r \right)}\left( g \right)
    \chi^{\left( s \right)}\left( {g^{-1}} \right)}\ =\ 0
\ , \quad \quad
    \hbox{ unless } \Delta k_{ij} = 0
\quad .
\label{A.6}
\ee
For the rest of this appendix,
we assume
\be
  k_{ij}^{\left( r \right)} = k_{ij}^{\left( s \right)}
\ , \quad \quad
    \hbox{ for all } i<j
\quad .
\label{A.7}
\ee
When $\Delta k_{ij} = 0$,
the sum over $n_{ij}$ produces a factor of
\be
  \left( \prod_{i<j} \sum\limits_{n_{ij}} \right) 1\ =\
    \prod_{i<j} {N_{ij}}
\quad .
\label{A.8}
\ee

Given that the $k_{ij}$ are
the same for both representations,
the following prime and derived quantities are equal
$$
  k_{ij}^{\prime \left( r \right)} =
  k_{ij}^{\prime \left( s \right)}
\ , \quad \quad
  N_{ij}^{\prime \left( r \right)} =
  N_{ij}^{\prime \left( s \right)}
\ , \quad \quad
    \hbox{ for } 1 \le i<j \le 4
\quad ,
$$
$$
  N_{\ell}^{\prime \left( r \right)} =
  N_{\ell}^{\prime \left( s \right)}
\ , \quad \quad
  M_{\ell}^{\left( r \right)} =
  M_{\ell}^{\left( s \right)}
\quad ,
$$
\be
  a_i^{\left( {\ell} \right)\left( r \right)} =
  a_i^{\left( {\ell} \right)\left( s \right)}
\ , \quad
  b_i^{\left( {\ell} \right)\left( r \right)} =
  b_i^{\left( {\ell} \right)\left( s \right)}
\ , \quad
  a_i^{\prime \left( {\ell} \right)\left( r \right)} =
  a_i^{\prime \left( {\ell} \right)\left( s \right)}
\ , \quad
  b_i^{\prime \left( {\ell} \right)\left( r \right)} =
  b_i^{\prime \left( {\ell} \right )\left( s \right)}
\quad ,
\label{A.9}
\ee
for $i=1,2,3,4$ and for all $\ell$.

Explicit examination of $P_{(N)}$ and $Q_{(N)}$
reveals that the traces
$\Tr P_{(N)}^n$ and  $\Tr Q_{(N)}^n$ are zero
unless $n = 0 \ ( \bmod {N} )$.
In prime-factorized form,
this holds for each $\ell$ sector.
Hence,
for $g$ in the form
of Eq.\ \bref{A.3},
$
 \chi^{\left( r \right)} \left( g \right) = 0
$
unless, for each $ \ell $,
$$
  \sum\limits_{i=1}^4 {a_i^{\left( {\ell} \right)}n_i} = 0
    \pmod{ p_{\ell}^{s_{\ell}} }
\quad ,
$$
$$
  \sum\limits_{i=1}^4 {b_i^{\left( {\ell} \right)}n_i} = 0
    \pmod{ p_{\ell}^{s_{\ell}} }
\quad ,
$$
$$
  \sum\limits_{i=1}^4 {a_i^{\prime \left( {\ell} \right)}n_i} = 0
    \pmod{ p_{\ell}^{u_{\ell}} }
\quad ,
$$
\be
  \sum\limits_{i=1}^4 {b_i^{\prime \left( {\ell} \right)}n_i} = 0
    \pmod{ p_{\ell}^{u_{\ell}} }
\quad .
\label{A.10}
\ee
The $4L$ constraints
in Eq.\ \bref{A.10}
need to be imposed on the $n_i$
in the sum
in Eq.\ \bref{A.2}.
When these constraints are satisfied,
\be
  \chi^{\left( r \right)}\left( g \right)
  \chi^{\left( s \right)}\left( {g^{-1}} \right)\ =\
    d_r^2
    \exp \left[ {
       2\pi i\sum\limits_{i=1}^4
       { { {\Delta k_i n_i} \over {L_i} } }
                 } \right]
\quad ,
\label{A.11}
\ee
where
$
  \Delta k_{i} \equiv
    k_{i}^{\left( r \right)} - k_{i}^{\left( s \right)}
$.

Summarizing the situation at this stage,
\be
  \sum\limits_{g\in G}
    {\chi^{\left( r \right)}\left( g \right)
     \chi^{\left( s \right)}\left( {g^{-1}} \right)}\ =\
  d_r^2\
  \delta_{k_{ij}^{\left( r \right)}, k_{ij}^{\left( s \right)}}
  \left( {\prod_{i<j} {N_{ij}}} \right) \!\!
   \sum\limits_{\scriptstyle {n_1,n_2,n_3,n_4}
 \atop \scriptstyle {with\ constraints} } \!\!\!
    \exp \left[ {
        2\pi i\sum\limits_{i=1}^4   {
      {{\Delta k_in_i} \over {L_i}} }
                } \right]
\ ,
\label{A.12}
\ee
where the constraints on the $n_i$
are given
in Eq.\ \bref{A.10}.

A trick can be used to handle the constraints.
To save space,
it is convenient to define
$$
  a_{1i}^{\left( {\ell} \right)} \equiv
    a_i^{\left( {\ell} \right)}
\ , \quad
  a_{2i}^{\left( {\ell} \right)} \equiv
    b_i^{\left( {\ell} \right)}
\ , \quad
  a_{3i}^{\left( {\ell} \right)} \equiv
    a_i^{\prime \left( {\ell} \right)}
\ , \quad
  a_{4i}^{\left( {\ell} \right)} \equiv
    b_i^{\prime \left( {\ell} \right)}
\quad ,
$$
\be
  s_{1\ell} \equiv s_{\ell}
\ , \quad
  s_{2\ell} \equiv s_{\ell}
\ , \quad
  s_{3\ell} \equiv u_{\ell}
\ , \quad
  s_{4\ell} \equiv u_{\ell}
\quad ,
\label{A.13}
\ee
so that the constraints
in Eq.\ \bref{A.10}
can be written in compact form as
\be
  \sum\limits_{i=1}^4 {a_{ci}^{\left( {\ell} \right)} n_i} =
   0 \pmod{ p_{\ell}^{s_{c\ell}} }
\ , \quad \quad
   c = 1, 2, 3, 4
\ , \quad
   \ell = 1, 2, \ldots , L
\quad .
\label{A.14}
\ee
The factor
\be
    \prod_{\ell=1}^L \prod_{c=1}^4 {}
  \left( {
    {1 \over {p_{\ell}^{s_{c\ell}}}}
  \sum\limits_{m_c^{\left( {\ell} \right)}=1}^{p_{\ell}^{s_{c\ell}}}
     \exp \left[ {
     -2\pi i{{m_c^{\left( {\ell} \right)}} \over
        {p_{\ell}^{s_{c\ell}}}}\sum\limits_{i=1}^4
     {a_{ci}^{\left( {\ell} \right)}n_i}
                 } \right]
          } \right)
\label{A.15}
\ee
automatically produces the constraints on the $n_i$
when the sums over the $m_c^{\left( {\ell} \right)}$
are performed.
In other words,
$$
  \sum\limits_{\scriptstyle {n_1,n_2,n_3,n_4} \atop
  \scriptstyle { with\ constraints } } =
  \left( {\prod_{i=1}^4 {}\sum\limits_{n_i=1}^{L_i} {}} \right)
   \left( {
     \prod_{\ell=1}^L {} \prod_{c=1}^4 {}
   \left( {
     {1 \over {p_{\ell}^{s_{c\ell}}}}
  \sum\limits_{m_c^{\left( {\ell} \right)}=1}^{p_{\ell}^{s_{c\ell}}}
   \exp \left[ {
        -2\pi i{{m_c^{\left( {\ell} \right)}}
          \over {p_{\ell}^{s_{c\ell}}}}
        \sum\limits_{i=1}^4
        {a_{ci}^{\left( {\ell} \right)} n_i}
               } \right]
           } \right)
           } \right)
\ .
$$
\be
{}
\label{A.16}
\ee
Hence, we use
Eq.\ \bref{A.16} in Eq.\ \bref{A.12}
and sum over the $n_i$ freely.
When the $n_i$ sums are done before
the $m_c^{\left( {\ell} \right)}$ sums,
one finds a non-zero result if and only if
\be
  \Delta k_i\
   =\ L_i \sum\limits_{\ell,c}
    {{{m_c^{\left( {\ell} \right)}
    a_{ci}^{\left( {\ell} \right)}
      } \over {p_{\ell}^{s_{c\ell}}}}}
      \pmod{ L_i }
\ , \quad \quad
\hbox{ for } i = 1, 2, 3, 4
\quad .
\label{A.17}
\ee
Using the explicit form of the
$a_{ci}^{\left( {\ell} \right)}$
in Eqs.\ \bref{3.32} and \bref{A.13},
one can check that two sets of
$m_c^{\left( {\ell} \right)}$
cannot lead to the same $\Delta k_i$.
This means that, at most, one term
in the $m_c^{\left( {\ell} \right)}$ sum contributes.
If the momenta
$ k_{i}^{\left( r \right)}$ and $k_{i}^{\left( s \right)}$
of two representations $r$ and $s$
differ in a way given
by Eq.\ \bref{A.17},
then they are related by the element
\be
\prod_{\ell = 1}^{L} {
  \left( {
           E_{P \times I}^{\left( \ell \right)}
         } \right)^{m_1^{ \left( {\ell} \right)}}
  \left( {
           E_{Q^{-1}\times I}^{\left( \ell \right)}
         } \right)^{m_2^{ \left( {\ell} \right)}}
  \left( {
           E_{I \times P}^{\left( \ell \right)}
         } \right)^{m_3^{ \left( {\ell} \right)}}
  \left( {
           E_{I \times Q^{-1}}^{\left( \ell \right)}
         } \right)^{m_4^{ \left( {\ell} \right)}}
                    }
\quad ,
\label{A.18}
\ee
of $E$,
as can be seen
from Eqs.\ \bref{3.36}--\bref{3.39}.
Consequently,
\be
  r \sim  s
\quad ,
\label{A.19}
\ee
if a non-zero result is to be obtained.
Summarizing,
when the representations $r$ and $s$
are not equivalent,
Eq.\ \bref{A.2} gives zero.

When $r \sim s$,
there is a unique non-zero term
in the $m_c^{\left( {\ell} \right)}$ sum,
and the phase factors
in Eqs.\ \bref{A.12} and \bref{A.16}
cancel.
Then the sums over the ${n_i}$
produce ${L_1 L_2 L_3 L_4}$
and the $1/p_{\ell}^{s_\ell}$ factors
in Eq.\ \bref{A.16}
yield
$  1 / { \prod_\ell p_{\ell}^{2 s_\ell + 2 u_\ell} } =
  {1 / {d_r^2}}
$.
Inserting these results
into Eq.\ \bref{A.12}
leads to
$$
  \sum\limits_{g\in G}
   {\chi^{\left( r \right)}\left( g \right)
    \chi^{\left( s \right)}\left( {g^{-1}} \right)}\ =\
     d_r^2\ \delta^{\left( r \right)\left( s \right)}
      {{L_1 L_2 L_3 L_4} \over {d_r^2}}
    \left( {\prod_{i<j} {N_{ij}}} \right)
$$
\be
    \phantom{.}\qquad\qquad\qquad
      =\ \delta^{\left( r \right)\left( s \right)}
    \left( { \prod_{i=1}^4 {L_i} }\right)
    \left( {\prod_{i<j} {N_{ij}}} \right)\ =\
    \delta^{\left( r \right) \left( s \right)}
    o\left( G \right)
\quad .
\label{A.20}
\ee

Equations \bref{A.1} and \bref{A.20} guarantee that
all the irreducible representations have been found.

\vfill\eject

\end{document}